\documentclass[aip,graphicx,amsmath,amssymb,reprint]{revtex4-1}

\usepackage{graphicx}
\usepackage{dcolumn}
\usepackage{bm}
\usepackage[colorlinks,breaklinks,linkcolor=blue,anchorcolor=blue,citecolor=blue,urlcolor=blue]{hyperref}
\usepackage[utf8]{inputenc}
\usepackage[T1]{fontenc}
\usepackage{mathptmx}
\usepackage{etoolbox}
\usepackage{amsfonts}
\usepackage{textcomp}
\usepackage{braket}
\usepackage{titlesec}

\usepackage[normalem]{ulem}
\begin{document}


\title{Rydberg superatoms: An artificial quantum system for quantum information processing and quantum optics} 



\author{Xiao-Qiang Shao}
\email[]{Authors to whom correspondence should be addressed: shaoxq644@nenu.edu.cn; slsu@zzu.edu.cn; li\_lin@hust.edu.cn; rejish@iiserpune.ac.in; jhwu@nenu.edu.cn; and weibin.li@nottingham.ac.uk }
\affiliation{Center for Quantum Sciences and School of Physics, Northeast Normal University, Changchun 130024, China}

\author{Shi-Lei Su}
\email[]{Authors to whom correspondence should be addressed: shaoxq644@nenu.edu.cn; slsu@zzu.edu.cn; li\_lin@hust.edu.cn; rejish@iiserpune.ac.in; jhwu@nenu.edu.cn; and weibin.li@nottingham.ac.uk }
\affiliation{School of Physics and Laboratory of Zhongyuan Light, Key Laboratory of Materials Physics of Ministry of Education, Zhengzhou University, Zhengzhou 450001, China
}%

\author{Lin Li}
\email[]{Authors to whom correspondence should be addressed: shaoxq644@nenu.edu.cn; slsu@zzu.edu.cn; li\_lin@hust.edu.cn; rejish@iiserpune.ac.in; jhwu@nenu.edu.cn; and weibin.li@nottingham.ac.uk }
\affiliation{MOE Key Laboratory of Fundamental Physical Quantities Measurement, Hubei Key Laboratory of Gravitation and Quantum Physics, PGMF, Institute for Quantum Science and Engineering, School of Physics, Huazhong University of Science and Technology, Wuhan 430074, China
}%

\author{Rejish Nath}
\email[]{Authors to whom correspondence should be addressed: shaoxq644@nenu.edu.cn; slsu@zzu.edu.cn; li\_lin@hust.edu.cn; rejish@iiserpune.ac.in; jhwu@nenu.edu.cn; and weibin.li@nottingham.ac.uk }
\affiliation{Indian Institute of Science Education and Research, Pune-411008, India
}%

\author{Jin-Hui Wu}
\email[]{Authors to whom correspondence should be addressed: shaoxq644@nenu.edu.cn; slsu@zzu.edu.cn; li\_lin@hust.edu.cn; rejish@iiserpune.ac.in; jhwu@nenu.edu.cn; and weibin.li@nottingham.ac.uk }
\affiliation{Center for Quantum Sciences and School of Physics, Northeast Normal University, Changchun 130024, China
}%

\author{Weibin Li}
\email[]{Authors to whom correspondence should be addressed: shaoxq644@nenu.edu.cn; slsu@zzu.edu.cn; li\_lin@hust.edu.cn; rejish@iiserpune.ac.in; jhwu@nenu.edu.cn; and weibin.li@nottingham.ac.uk }
\affiliation{School of Physics and Astronomy, and Centre for the Mathematics and Theoretical Physics of Quantum Non-equilibrium Systems, The University of Nottingham, Nottingham NG7 2RD, United Kingdom
}%

\date{\today}

\begin{abstract}
Dense atom ensembles with Rydberg excitations display intriguing collective effects mediated by their strong, long-range dipole-dipole interactions. These collective effects, often modeled using Rydberg superatoms, have gained significant attention across various fields due to their potential applications in quantum information processing and quantum optics. In this review article, we delve into the theoretical foundations of Rydberg interactions and explore experimental techniques for their manipulation and detection. We also discuss the latest advancements in harnessing Rydberg collective effects for quantum computation and optical quantum technologies. By synthesizing insights from theoretical studies and experimental demonstrations, we aim to provide a comprehensive overview of this rapidly evolving field and its potential impact on the future of quantum technologies. 
\end{abstract}


\maketitle 

\tableofcontents

\section{INTRODUCTION}


Rydberg atoms exhibit striking properties that arise from their { large} atomic size and strong interatomic interactions.\cite{TFGallagher_1988,gallagher1994rydberg,RevModPhys.82.2313} {The huge induced electric dipole moment} couples with external electric fields, allowing the manipulation of the electronic as well as motional states of Rydberg atoms with laser and microwave (MW) fields. {Importantly}, {strong and long-range dipole-dipole or van der Waals (vdW) interactions are found between Rydberg atoms separated by macroscopic distances of over 10 micrometers}.\cite{urban2009observation,gaetan2009observation} Such interactions can be controlled by laser or MW fields. This paves the way for realizing scalable quantum information processing architectures,~\cite{PhysRevLett.85.2208} creating novel many-body states~\cite{browaeys2020many} and exploring nonlinear quantum optics.\cite{Firstenberg_2016} 

{At the heart of the Rydberg physics lies strong interaction-induced collective behaviors.\cite{PhysRevLett.99.163601,dudin2012observation,PhysRevX.5.031015}} 
In high-lying Rydberg states, { the interatomic interaction ($1$--$100$~MHz) can be much stronger than typical laser Rabi frequencies (tens of kHz--a few~MHz).\cite{RevModPhys.82.2313}} As a result, the laser excitation of one atom inhibits the excitation of neighboring atoms within a certain distance (i.e. blockade radius).
Within the blockade radius, {only the many-body ground state and singly excited Rydberg state can be occupied.} The former consists of atoms in the electronic ground state, while the latter is a collective state where one atom is in the Rydberg state while all other atoms are in the electronic ground state. Dynamics of such collective states are conveniently described by an artificial \textit{superatom}.\cite{PhysRevLett.87.037901} This description can be extended to many-atom settings, where the collective effect allows for the creation of exotic many-body correlated states.\cite{PhysRevLett.104.043002,PhysRevLett.103.185302,PhysRevLett.105.160404,weimer2010rydberg,PhysRevLett.107.060406,de2019observation} These potentials make Rydberg atoms a leading candidate for implementing quantum simulation and quantum computing tasks. Moreover, strong Rydberg atom interactions can be mapped to optical fields, generating sizable optical nonlinearities.\cite{PhysRevLett.105.193603} These have found wide usage in exploring nonlinear quantum optics~\cite{peyronel2012quantum, Firstenberg_2016} and realizing optical quantum devices, such as single-photon sources,\cite{doi:10.1126/science.1217901,ripka2018room,PhysRevResearch.3.033287,Ornelas-Huerta:20, Shi2022} photonic quantum logic gates,\cite{doi:10.1126/sciadv.1600036,Tiarks2019,PhysRevX.12.021035} photonic entanglement filter,\cite{ye2023photonic} single-photon absorber,\cite{PhysRevLett.117.223001} single-photon 
switch,\cite{PhysRevLett.112.073901} and single-photon transistors.\cite{PhysRevLett.113.053601}

{In recent years, {quite a few} review papers have meticulously surveyed the rapidly evolving field of Rydberg atom technologies,\cite{RevModPhys.82.2313,browaeys2020many,Saffman_2016,Schauss_2018,Adams_2020,Beterov_2020,Henriet2020quantumcomputing,10.1116/5.0036562,Wu_2021,Shi_2022,Yuan_2023} emphasizing their diverse applications and groundbreaking advancements in quantum computation, quantum simulation, and quantum sensing, showcasing the versatility and potential of Rydberg atoms. This review article distinguishes itself by concentrating specifically on the concept of Rydberg superatoms, with a particular emphasis on their role as a quantum interface between atoms and photons.
Rydberg superatoms are created through collective excitation states, where multiple atoms collectively display characteristics akin to a single two-level atom. This collective behavior not only enhances the coherence and controllability of quantum states but also endows the interface with novel electromagnetic properties, enabling the creation and manipulation of photon-photon interactions at the quantum level. This review thus will contribute to the existing body of knowledge on Rydberg research, offering valuable insights and guidance for future studies.}

{In this review article, we delve into the theory underpinning the Rydberg superatom, related collective effects, and key experimental achievements. In particular, we review the {harnessing} of Rydberg collectivity in two critical areas: quantum information processing and quantum optics. We start with an introduction to Rydberg atoms and Rydberg superatoms in Sec.~\ref{sec:Rydbergatoms}, discussing the atomic properties, two-body interactions of Rydberg atoms, and the emergence of collective states described by Rydberg superatoms. In Sec.~\ref{sec:experimentaltech}, coherent laser excitation of Rydberg atoms and the preparation of Rydberg superatoms {are covered}, including quantum interference of Rydberg superatoms. In Sec.~\ref{sec:quantuminformation}, we discuss quantum information processing based on Rydberg superatoms, focusing on implementing quantum logic gates, generating quantum entanglement, and performing quantum simulations. We describe the application of Rydberg superatoms in quantum optics in Sec.~\ref{sec:quantumoptics}, including the demonstration and realization of strong nonlocal nonlinear optics, single-photon applications, and hybridizing quantum systems with Rydberg superatoms. Finally, we conclude in Sec.~\ref{concl}.}

\section{RYDBERG ATOMS AND RYDBERG SUPERATOMS}
\label{sec:Rydbergatoms}

\subsection{Characteristics of Rydberg atoms}

In a broad sense, a Rydberg atom can be defined as an atom with one of its valence electrons in a highly excited state characterized by a large principal quantum number. The category of ``atoms'' that exhibit this characteristic includes alkali atoms,~\cite{TFGallagher_1988,RevModPhys.82.2313,lim_review_2013} alkaline earth atoms,\cite{PhysRevLett.121.203001,PhysRevX.8.041055,zhang2020controlling,schafer2020tools,kaufman2021quantum,norcia2021developments,Mukherjee_2011,PhysRevLett.128.033201,PhysRevX.12.011054,muni2022optical,wu2022erasure} trapped ions,\cite{Müller_2008,PhysRevLett.115.173001,PhysRevLett.108.023003,PhysRevLett.119.220501,PhysRevA.100.022513,zhang2020submicrosecond,PhysRevLett.126.233404} as well as solid state systems.\cite{kazimierczuk2014giant,https://doi.org/10.1002/qute.201900134,PhysRevB.96.125142}
Since alkali atoms possess only one valence electron, their properties in the Rydberg state closely resemble those of a hydrogen atom. This allows us to derive the distinctive properties of alkali Rydberg atoms using, for example, the quantum defect theory.~\cite{TFGallagher_1988,gallagher1994rydberg}
The binding energy of the Rydberg states of alkali atoms (e.g., Rb and Cs) with the principal quantum number $n$, orbital angular momentum $l$, and total angular momentum $j=l\pm1/2$ can be expressed as
\begin{equation}
E_{n,l,j}=-\frac{hc R_{m}}{(n-\delta_{n,l,j})^2}.    
\end{equation}
Here, $R_{m}=R_{\infty}/(1+m_e/m_{\rm {nucleus}})$ represents the mass-corrected
Rydberg constant, where $m_e$ denotes the electron mass and
$m_{\rm {nucleus}}$ refers to the atomic mass of the nucleus. The term $R_{\infty}=m_ee^4/(8\epsilon_0^2h^3c)$ is the Rydberg constant, where $e$ is the elementary charge, $\epsilon_0$ denotes the permittivity of the vacuum, $h$ represents the Planck constant, $c$ stands for the speed of light in vacuum.\cite{MJSeaton_1983,gallagher1994rydberg} Furthermore, the species-dependent quantum defect, denoted as $\delta_{n,l,j}$, can be approximated by the modified Rydberg-Ritz parameters 
\begin{equation}
\delta_{n,l,j}=\delta_0+\frac{\delta_2}{(n-\delta_0)^2}+\frac{\delta_4}{(n-\delta_0)^4}+\frac{\delta_6}{(n-\delta_0)^6}+...,    
\end{equation}
where the coefficients $\delta_0$, $\delta_2$, $\delta_4$, and $\delta_6$ are primarily depends on $l$ (while also being related to $j$) and remains non-negligible solely for states with $l\leq3$.\cite{C-JLorenzen_1983,gallagher1994rydberg}
Precise understanding of the quantum defect is paramount for anticipating spectral lines, comprehending electron behavior in highly excited states, and probing the boundaries of quantum mechanics. This imperative has led researchers to pursue quantum defect parameters of Rydberg atoms with high precision through various methodologies.\cite{PhysRevA.26.2733,lorenzen1984precise,PhysRevA.35.4650,TFGallagher_1988,PhysRevA.52.514,PhysRevA.67.052502,PhysRevA.74.054502,PhysRevA.74.062712,Sanguinetti_2009,PhysRevA.83.052515,PhysRevA.93.013424,Sautenkov_2016,Li:19,LI2021104728,bai2023quantum} In Table \ref{defect}, we present a comprehensive compilation of modified Rydberg-Ritz parameters for low orbit states in Rb and Cs atoms. This compilation incorporates data from early seminal work and recent measurements. For additional orbitals or other alkali atoms, their quantum defects can be easily obtained using two recently released open-source software packages, pairinteraction\cite{Weber_2017} and Alkali
Rydberg Calculator (ARC) toolbox.\cite{SIBALIC2017319}

\begin{table}
\centering
\caption{\label{defect}Quantum defect constants $\delta_0$, $\delta_2$, $\delta_4$, and $\delta_6$ for different orbital
angular momentum $l$ of Rb and Cs extracted from Refs.~\onlinecite{C-JLorenzen_1983,lorenzen1984precise, PhysRevA.67.052502, PhysRevA.93.013424,LI2021104728,bai2023quantum}.}
\setlength{\tabcolsep}{1.0mm}
\begin{ruledtabular}
\begin{tabular}{cccccc}
Rydberg series&$\delta_0$&$\delta_2$&$\delta_4$&$\delta_6$\\
\hline
$^{85}$Rb~$nS_{1/2}$&3.1311804\cite{PhysRevA.67.052502}&0.1784\cite{PhysRevA.67.052502}&-1.8\cite{C-JLorenzen_1983} & \\
$^{85}$Rb~$nP_{3/2}$& 2.64142\cite{LI2021104728}& 0.295\cite{LI2021104728}& 0.97495\cite{C-JLorenzen_1983} & 14.6001\cite{C-JLorenzen_1983} \\
$^{85}$Rb~$nD_{5/2}$&1.3464572\cite{PhysRevA.67.052502}&-0.596\cite{PhysRevA.67.052502}&-1.50517\cite{C-JLorenzen_1983} &-2.4206\cite{C-JLorenzen_1983} \\
$^{85}$Rb~$nF_{7/2}$& 0.016411\cite{LI2021104728}&-0.0784\cite{LI2021104728} &-0.36005\cite{C-JLorenzen_1983} & 3.2390\cite{C-JLorenzen_1983}\\
\hline
$^{133}$Cs~$nS_{1/2}$& 
4.0493532\cite{PhysRevA.93.013424}& 0.23915\cite{PhysRevA.93.013424}& 0.06\cite{PhysRevA.93.013424}&11\cite{PhysRevA.93.013424}\\
$^{133}$Cs~$nP_{3/2}$&3.5590676\cite{PhysRevA.93.013424}&  0.37469\cite{PhysRevA.93.013424}&-0.67431\cite{lorenzen1984precise}
&22.3531\cite{lorenzen1984precise}\\
$^{133}$Cs~$nD_{5/2}$&2.4663144\cite{PhysRevA.93.013424}& 0.01381\cite{PhysRevA.93.013424}&-0.392\cite{PhysRevA.93.013424}&-1.9\cite{PhysRevA.93.013424}\\
$^{133}$Cs~$nF_{7/2}$& 0.0335646\cite{bai2023quantum}&-0.2052\cite{bai2023quantum}& & \\
\end{tabular}
\end{ruledtabular}
\end{table}

Starting from the expression for Rydberg's binding energy, we can obtain a series of scaling laws that exhibit {outstanding} characteristics concerning the principal quantum number $n$, or more precisely, the effective principal quantum number $n^*=(n-\delta_{n,l,j})$. The energy spacing between adjacent Rydberg levels is given by
\begin{equation}
\Delta E=E_{n+1,l+1,j}-E_{n,l,j}\approx\frac{2hcR_{m}[(\delta_{n,l,j}-\delta_{n,l+1,j})+1]}{n^{*3}},
\end{equation}
which follows a $n^{*-3}$ scaling. The corresponding transition wavelength typically falls within the millimeter-wave band between 10~mm (30~GHz) and 1~mm (300~GHz).\cite{SIBALIC2017319} {This range of wavelengths} offers exceptional precision in spectroscopic measurements, \cite{PhysRevLett.107.143001,10.1063/1.4890094} showing transient amplifiers of microwave radiation \cite{PhysRevLett.49.1924,PhysRevA.27.2043,PhysRevA.27.2065}  and contributing to the development of the fifth generation mobile communication network.\cite{10.1063/1.5031033,10189894} It serves a crucial role in the advancement of atomic physics and quantum science research.\cite{PhysRevA.88.043429,kiffner2016two,PhysRevA.95.043818,wade2017real,PhysRevA.100.032512,10.1063/1.5137900,kanungo2022realizing,legaie2023millimeter}

\begin{figure}
\centering  
\includegraphics[width=1\linewidth]{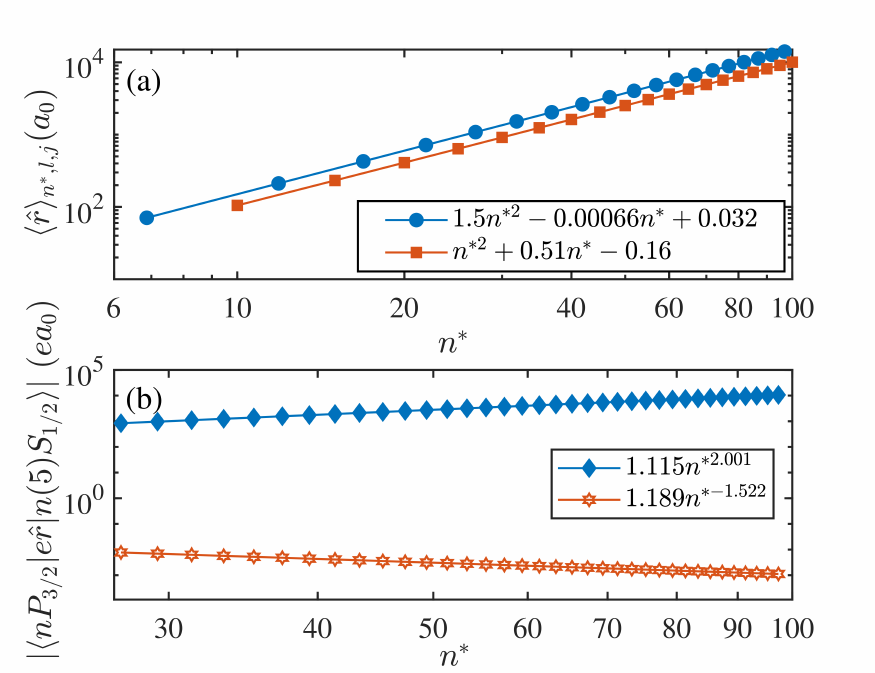}
\caption{\label{scaling} (a) The numerical data demonstrates the scaling laws for the average orbit radius of Rydberg atoms in the $S$ orbit {(blue circles) and circular orbit (red squares)}, expressed in terms of the effective principal quantum number denoted as $n^*$ for each case. (b) The scaling laws govern the radial dipole matrix elements for transitions between neighboring Rydberg states, denoted as $|\langle nP_{3/2}|e\hat{r}|nS_{1/2}\rangle|$ {(blue diamonds)}, as well as for transitions between a low-lying state and a Rydberg state, represented as $|\langle nP_{3/2}|e\hat{r}|5S_{1/2}\rangle|$ {(red hexagons)}, where $n^* = n - \delta_{n,2,j}$ for the $P$ orbit.}
\end{figure}

The {expectation value} of the electron orbit, determined by the radial wave function {for an  alkali atom state with medium to high quantum numbers $n$, }is given by\cite{LINDGARD1977533,PhysRevA.30.2881}
\begin{equation}
\langle \hat{r}\rangle_{n^*,l,j}=\frac{1}{2}[3n^{*2}-l(l+1)]a_0,
\end{equation}
where $a_0\approx5.29\times10^{-2}~\rm~nm$ represents the Bohr radius. This formula offers an exact description of the radius of Rydberg atomic orbitals with high principal quantum numbers, as the Coulomb potential predominates in this regime. {The Rydberg atomic wave function is influenced by quantum defects. They mainly generate a phase shift to the wave function. Hence wave functions of alkali atoms are different from that of hydrogen atoms as long as quantum defects are non-negligible.} As a result, the {expectation value} of physical quantities can be directly derived from the {hydrogen wave function}, with the only alteration being the substitution of the principal quantum number $n$ with the effective principal quantum number $n^*$.\cite{kazimierczuk2014giant,PhysRevB.96.125142}
For low $l$ states, particularly $l=0$ ($S$ states), the {expectation value} simplifies to be $3n^{*2}a_0/2$, but for the circular orbit with $l=n-1$, where the quantum defect is zero, the value becomes $(n^2+n/2)a_0$. For large $n$ the radius of the Rydberg electron can be hundreds of nanometers. This permits the formation of ultra-long range Rydberg molecules,~\cite{doi:10.1146/annurev.pc.38.100187.000331,PhysRevLett.85.2458,bendkowsky2009observation,PhysRevLett.105.163201,PhysRevA.86.031401,PhysRevLett.112.143008,PhysRevA.93.022702,liHomonuclearMoleculePermanent2011,PhysRevA.95.042515,PhysRevA.95.052708,shaffer2018ultracold,PhysRevA.101.060701,PhysRevA.102.033315,PhysRevLett.126.013001,doi:10.1080/00268976.2019.1679401,PhysRevLett.127.023003} where ground state atoms are trapped within the Rydberg electron cloud. 
In Fig.~\ref{scaling}(a), the ARC toolbox \cite{SIBALIC2017319} is used to perform numerical calculation of the {expectation values} for the $S$ and circular orbits of Rubidium, encompassing the principal quantum number $n$ from 10 to 100. The data exhibit quadratic scaling with $n^*$, closely aligning with the analytical expression.

Electric dipole transitions $\langle n',l',j'|e\hat{r}|{n},l,j\rangle$ between Rydberg atomic states are also characterized by elements of the radial matrix $\int R_{n',l',j'}(r)erR_{n,l,j}(r)r^2dr$ together with the angular components.\cite{Sobelman} These elements show a scaling of $n^{*2}$ for adjacent Rydberg states (${{n}'}\approx n$), consistent with the behavior of the orbital radius. Moreover, it scales with $n^{*-1.5}$ for transitions between a low-lying state and a Rydberg state (${n}'\ll n$),  due
to the smaller overlap of the radial wave function, where low-lying states localize around the ion core.\cite{TFGallagher_1988,DEIGLMAYR2006293} 
These scaling laws find an intuitive representation in Fig.~\ref{scaling}(b), where we have specifically chosen the elements of the dipole radial matrix $|\langle nP_{3/2}|e\hat{r}|nS_{1/2}\rangle|$ and $|\langle nP_{3/2}|e\hat{r}|5S_{1/2}\rangle|$ as illustrative examples. The other properties of the Rydberg states, such as their spontaneous decay rate and response to electric fields, are closely related to these dipole matrix elements.

{Lifetimes of Rydberg atoms are much longer than typical time scales of laser coupling and Rydberg atom interactions. This is important for maintaining quantum coherence, a crucial factor for quantum computation and simulation. The spontaneous decay rate from $|n,l,j\rangle$ to lower energy state $|n',l',j'\rangle$ is governed by the Einstein coefficient (in SI units):
\begin{eqnarray}
A_{nlj\rightarrow n'l'j'}&=&\frac{2\omega^3_{n'l'j',nlj}}{3\epsilon_0 h c^3}(2j'+1)l_{\rm max}
\left\{
\begin{array}{ccc}
l,&l',&1\\
j',&j,&1/2
\end{array}\right\}^2\nonumber\\
&&\times |\langle n',l',j'|e\hat{r}|{n},l,j\rangle|^2,
\end{eqnarray}
where  $\omega_{n'l'j',nlj}$ denotes the corresponding transition angular frequency,  
$l_{max}$ signifies the larger of the two values between $l$ and $l'$, and the curly braces indicate a Wigner-$6j$ symbol.\cite{Seiler_2016,PhysRevA.79.052504} {The radial part of Rydberg electronic wave functions has a vanishing overlap with that of ground and weakly excited states, decreasing $A_{nlj\rightarrow n'l'j'}$.  At zero temperature, the overall decay rate of Rydberg states to many low-lying states is proportional to $n^{*-3}$, where in this case $\omega_{n'l'j',nlj}$ is nearly constant and only the electric dipole transitions dominate. However, in circular Rydberg states ($l=n-1$), because $A_{nlj\rightarrow n'l'j'}\propto \omega^3_{n'l'j',nlj}$, they can only undergo a transition to an adjacent Rydberg state via a dipole-allowed process.\cite{PhysRevLett.51.1430,PhysRevX.8.011032,PhysRevX.14.021024} This leads to the spontaneous emission rate of the circular state being proportional to $n^{*-5}$.}}

Compared to the high $l$ states, the quantum defects of the low $l\leq3$ states have a greater impact on the energy levels correction. Consequently, the weak external { static} electric field will not cause the Rydberg atom in the low orbit state to endure a direct transition between the adjacent orbits but will instead present the energy level shift as a second-order perturbation, which is in the form of a quadratic Stark effect\cite{PhysRev.167.128,angel1968hyperfine}
\begin{equation}
\Delta E_{\rm Stark}=\sum_{n',l'j'\neq n,l,j}\frac{|\langle n',l',j',m_j'|e\hat{r}\cdot \boldsymbol{E}|n,l,j,m_j
\rangle|^2}{E_{n',l',j'}-E_{n,l,j}}=-
\frac{1}{2}\alpha_0 E^2
\end{equation}
{where we have assumed the static electric field $\boldsymbol{E}$} is applied along the $z$-axis creating off-diagonal couplings between states, with the
selection rule $\Delta l=\pm1$, $\Delta m_j=0$. As determined by the Wigner-Eckart theorem,\cite{weissbluthatoms} the dipole matrix elements are linked to the radial matrix elements $\langle n',l',j'|e\hat{r}|{n},l,j\rangle$ through coefficients independent of the principal quantum number. Therefore, the scalar polarizability $\alpha_0$ scales with $n^{*7}$ due to the combined scaling of the radial dipole matrix element and the energy spacing for neighboring Rydberg states in the denominator.\cite{PhysRevA.20.2251,PhysRevA.31.2718,PhysRevA.98.052503,Bai_2020,PhysRevA.108.042818}
It should be noted that states possessing quantum numbers of angular momentum $l>3$ manifest degeneracy, leading to a linear Stark shift. For a more precise description of the Stark structure of the Rydberg states in alkali atoms, direct diagonalization of the energy matrix is required. The Rydberg atoms {due to high polarizability can be exploited as highly accurate sensors} for electric fields\cite{sedlacek2012microwave,Fan:14,Fan_2015,10.1063/1.5038550,PhysRevLett.121.110502,Meyer_2020,jing2020atomic,https://doi.org/10.1002/qute.202100049,PhysRevA.104.043103,PhysRevApplied.15.014047,10.1063/5.0086357,PhysRevApplied.18.014045,10.1063/5.0137127,10238372,dingEnhancedMetrologyCritical2022,Yuan_2023} and the nondestructive imaging of atoms and ions.\cite{PhysRevLett.108.013002,PhysRevLett.124.053401}

\begin{table}
\centering
\caption{\label{nscal}Scaling of various properties of Rydberg atoms with the effective principal quantum numbers $n^*$. The data for { the transition rate} in the parentheses is scaling in the circular state.}
\setlength{\tabcolsep}{0.2mm}
\begin{ruledtabular}
\begin{tabular}{ccc}
Property&Expression&$n^*$~scaling\\
\hline
Binding energy&$E_{n,l,j}=-\frac{{hcR_m}}{(n-\delta_{n,l,j})^2}$&$n^{*-2}$\\
Level spacing& $E_{n+1,l+1,j}-E_{n,l,j}$& $n^{*-3}$\\
Orbit radius&$\langle \hat{r}\rangle_{n^*,l,j}=\frac{1}{2}[3n^{*2}-l(l+1)]a_0$&$n^{*2} $\\
Dipole moment& $\langle nP_{3/2}|e\hat{r}|n(5)S_{1/2}\rangle$ & $n^{*2}$ ($n^{*-1.5}$) \\
Transition rate& $A_{nlj\rightarrow n'l'j'}$ &$n^{*-3} (n^{*-5})$ \\
 Scalar polarizability& $\alpha_0$&$n^{*7}$ \\
\end{tabular}
\end{ruledtabular}
\end{table}

In Table~\ref{nscal}, a concise overview is presented, detailing the scaling behavior of various properties in Rydberg atoms with respect to the effective principal quantum number. A comprehensive exploration of additional Rydberg atom properties and associated mechanisms can be found in numerous recent review articles and references
therein.\cite{RevModPhys.82.2313,Löw_2012,Browaeys_2016,Saffman_2016,Schauss_2018,Wüster_2018,Adams_2020,Beterov_2020,browaeys2020many,Henriet2020quantumcomputing,10.1116/5.0036562,Wu_2021,Shi_2022}

\subsection{Interaction potential between Rydberg atoms}

The interaction potential between Rydberg atoms manifests as an electrostatic interaction, which can be expressed as an infinite series of powers involving the inverse of the interatomic distance $R$.\cite{DALGARNO19661} The predominant component is represented by the lowest-order dipole terms\cite{PhysRevLett.85.2208,PhysRevA.62.052302}
\begin{eqnarray}
\hat{V}_{int}=\frac{1}{4\pi\epsilon_0}[\frac{\hat{{d_1}}\cdot\hat{{d_2}}-3(\hat{{d_1}}\cdot\boldsymbol{n})(\hat{{d_2}}\cdot\boldsymbol{n})}{R^3}],
\end{eqnarray}
where $\hat{d_i}=-e\hat{r_i}$ is the electric dipole operator of atom $i$ ($i=1,2$), and $\boldsymbol{n}$ denotes the unit vector connecting
the two atoms. Without loss of generality, the polar angle between the quantization axis $z$ and $\boldsymbol{n}$ is set to $\theta$, and the azimuthal angle is $\phi$. In the spherical basis, the electric dipole-dipole interaction can be expressed as
\begin{eqnarray}\label{vint}
\hat{V}_{int}&=&\frac{1}{4\pi\epsilon_{0}R^{3}}[\frac{1-3 \cos^{2}\theta}{2}(\hat{d}_{1}^+\hat{d}_{2}^-+\hat{d}_{1}^- \hat{d}_{2}^++2 \hat{d}_{1}^0\hat{d}_{2}^0)\nonumber\\
&&+\frac{3}{\sqrt{2}} \sin \theta \cos \theta(\hat{d}_{1}^+ \hat{d}_{2}^0e^{i\phi}-\hat{d}_{1}^- \hat{d}_{2}^0e^{-i\phi}+\hat{d}_{1}^0 \hat{d}_{2}^+e^{i\phi}\nonumber\\
&&-\hat{d}_{1}^0 \hat{d}_{2}^-e^{-i\phi})-\frac{3}{2} \sin ^{2} \theta(\hat{d}_{1}^+\hat{d}_{2}^+e^{2i\phi}+\hat{d}_{1}^-\hat{d}_{2}^-e^{-2i\phi})],
\end{eqnarray}
where $\hat{d}^{0}=\hat{d}^{z}$, $\hat{d}^{+}=-(\hat{d}^{x}+i \hat{d}^{y}) / \sqrt{2}$, and $\hat{d}^{-}=(\hat{d}^{x}-i\hat{d}^{y}) / \sqrt{2}$, which correspond to changes in  magnetic quantum number of $\Delta m_j=0$, $+1$, and $-1$, respectively, in the process of atomic transition. This formula enables us to construct typical interaction forms for the Rydberg atom system.  In the following, we will discuss different interaction types encountered in Rydberg atom experiments.

The spin-exchange interaction naturally occurs between two atoms in the quantum states of adjacent orbitals. As exemplified in Ref.~\onlinecite{PhysRevLett.114.113002}, considering a pair of Rydberg states for $^{87}$Rb atoms, denoted as $|\uparrow\rangle=|62D_{3/2},m_j=3/2\rangle$ and $|\downarrow\rangle=|63P_{1/2},m_j=1/2\rangle$, in the basis of two atoms $\{|\uparrow \uparrow\rangle,$ $|\uparrow \downarrow\rangle,$ $|\downarrow \uparrow\rangle,$ $|\downarrow \downarrow\rangle\}$, only the first line of Eq.~(\ref{vint}) is applicable due to the selection rules that govern the dipole operator. Consequently, the interaction Hamiltonian is,
\begin{equation}\label{spin}
\hat{H}_{\rm spin}=\frac{(1-3 \cos^{2}\theta)}{2}\frac{C_3}{R^3}(|\uparrow\downarrow\rangle\langle \downarrow\uparrow|+{\rm H.c.}),
\end{equation}
where ${\rm H.c.}$ stands for the Hermitian conjugate and the real coefficient $C_3=\langle\uparrow \downarrow\left|\hat{d}_{1}^+\hat{d}_{2}^-| \downarrow \uparrow\right\rangle / (4\pi\epsilon_0)$. 
This Hamiltonian was initially employed to examine the coherent dynamics of a spin excitation in a chain comprising three Rydberg atoms with $\theta=0$. \cite{PhysRevLett.114.113002,PhysRevLett.119.053202} Subsequently, it was extended to simulate a topological model protected by symmetry in a zigzag configuration of Rydberg atoms, with the magic angle {$\theta = \arccos (1/\sqrt{3})$} governing the hopping direction along the next nearest neighbor atoms. \cite{de2019observation} Additionally, it was utilized to synthesize a density-dependent Peierls phase employing three Rydberg atoms arranged in an equilateral configuration with $\theta=\pi/2$. \cite{PhysRevX.10.021031} Furthermore, this model finds application in the simulation of energy transport\cite{PhysRevLett.114.123005} and various generic spin chain models.\cite{geier2021floquet,PRXQuantum.3.020303,chen2023continuous,nishad2023quantum,chen2023spectroscopy,bornet2024enhancing,PhysRevA.109.042604}

{When the electronic energy of a pair of Rydberg atoms is (nearly) degenerate with energies of other Rydberg pair states, the dipole-dipole interaction can be strong due to the so-called F\"oster resonance.} Even if a small energy defect exists, it can be mitigated by the differential Stark effect induced by external electric fields. \cite{Walker_2005,PhysRevLett.104.073003} The mathematical expression closely resembles Eq.~(\ref{spin}), differing only in the composition of the bases. Experimental observations indicate that in the presence of a modest electric field ($F_{\rm res}\approx 33$~mV/cm), resonance occurs between the state $|dd\rangle$ and the symmetric state $|pf_1\rangle_{\rm s}=(|pf_1\rangle+|f_1p\rangle)/\sqrt{2}$ of two $^{87}$ Rb atoms, \cite{ravets2014coherent,PhysRevA.92.020701} where $|d\rangle=|59D_{3/2},m_j=3/2\rangle$, $|p\rangle=|61P_{1/2},m_j=1/2\rangle$ and $|f_1\rangle=|57F_{5/2},m_j=5/2\rangle$.
 Therefore, this interaction channel is effectively isolated from other interaction channels at an interatomic distance $R=9.1~\mu\rm m$ in the presence of a 3.3 G magnetic field. The corresponding Hamiltonian is given by
\begin{equation}\label{forster}
\hat{H}_{\rm F\ddot{o}rster}=\frac{(1-3 \cos^{2}\theta)}{\sqrt{2}}\frac{C_3}{R^3}(|dd\rangle_{\rm s}\langle pf_1|+{\rm H.c.}),
\end{equation}
with the measured $C_3/h = 2.39\pm0.03~{\rm GHz}~{\mu}{\rm m}^3$.
A more systematic study of F{\"o}rster resonance involving different atomic species can be found in  Ref.~\onlinecite{PhysRevA.92.042710}, and the Stark-tuned three-body F{\"o}rster resonances have been predicted \cite{faoro2015borromean} and observed.\cite{PhysRevLett.119.173402,PhysRevA.98.052703}
By integrating adiabatic rapid passage and Landau-Zener control with F{\"o}rster resonance tuning, robust two-qubit quantum logic gates can be achieved. \cite{PhysRevA.94.062307,PhysRevA.97.032701,PhysRevA.98.052324} Additionally, a recent experimental breakthrough has demonstrated an ultrafast F{\"o}rster oscillation capable of performing a conditional phase gate on nanosecond
timescales. \cite{chew2022ultrafast}

In the absence of an external electric field, the energy difference defined by $\Delta_{r',r''}=E_{r'}+E_{r''}-2E_r$ of the Rydberg atom pair in the F{\"o}rster resonance process greatly exceeds the strength of the dipole-dipole interaction when the atoms are at a significant distance from each other (perturbation theory is applicable). Here, $E_r$ represents the energy of the Rydberg state $|r\rangle$ of interest, while $E_{r'}$ and $E_{r''}$ denote the energies of a pair of Rydberg states $|r'\rangle$ and $|r''\rangle$, whose symmetric state is dipole coupled to the Rydberg pair state $|rr\rangle$. This leads to a multichannel-induced Stark shift in the energy levels of the atom pair, resulting in the interaction between the atoms manifesting as a classical vdw interaction,\cite{PhysRevA.77.032723,RevModPhys.82.2313} represented by 
\begin{equation}\label{vdw}
\hat{H}_{\rm vdw}=-\frac{C_6}{R^6}|rr\rangle\langle rr|.
\end{equation}
Here, $C_6$ is defined as
$C_6=\sum_{r'r''}|\langle r'r''|\hat{V}_{ int}|rr\rangle|^2/{\Delta_{r',r''}}$, where in principle $|r'r''\rangle$ encompasses all pair states dipole coupled to $|rr\rangle$. However, for practical computational efficiency, focusing on quantum states within a specific range of energy differences allows one to obtain sufficiently {accurate values of $C_6$}.
When the interatomic distance is not significantly large and perturbation theory is not applicable, it becomes crucial to perform the diagonalization of the Hamiltonian governing the pair-state interaction.
Experimentally, {the $|C_6|$ coefficients} for $|nD_{3/2},m_j=3/2\rangle$ of the $^{87}$Rb atom with $n=53$, 62, and 82 {are} measured directly between two isolated single Rydberg atoms separated by a controlled distance of a few micrometers,\cite{PhysRevLett.110.263201} and the angular dependence of the effective
interaction energy of the two $82D_{3/2}$ ($82S_{1/2}$) atoms is also demonstrated.\cite{PhysRevLett.112.183002} Recently,
 with the advancement of experimental techniques enabling the precise arrangement of a large number of atoms in defect-free programmable geometric configurations across various dimensions, the vdw-type Rydberg interaction has found applications in simulating large-scale Ising-type quantum spin models,\cite{bernien2017probing,PhysRevX.8.021070,PhysRevLett.120.113602,keesling2019quantum,ebadi2021quantum,scholl2021quantum,PhysRevLett.131.123201,PhysRevX.14.011025} generating and manipulating Schr{\"o}dinger cat states,\cite{doi:10.1126/science.aax9743} designing frustrated magnets,\cite{PhysRevX.4.041037, PhysRevLett.114.173002} and exploring nonequilibrium physics.\cite{ding_phase_2020,doi:10.1126/science.abg2530,PhysRevResearch.5.023010,PhysRevApplied.20.014014,10.1063/5.0192602}

\subsection{{Emergence of Rydberg superatoms}}

In the fields of chemistry and atomic physics, the concept of a superatom is used to characterize a cluster of atoms that exhibits collective behavior and possesses characteristics distinct from its constituents.\cite{PhysRevLett.52.2141,PhysRevLett.57.2560,PhysRevB.51.13705} These superatoms possess unique electronic structures and reactivity,\cite{luo2014special,reber2017superatoms} making them highly compelling for a wide range of applications in materials science and nanotechnology.\cite{castleman2009clusters,doud2020superatoms,yu2023atomic}
Similarly, a Rydberg superatom denotes a collectively excited state of an ensemble of $N$ atoms, {where the Rydberg blockade inhibits the simultaneous excitation of multiple Rydberg atoms in the ensemble.}\cite{PhysRevLett.85.2208} In a two-body system, this blockade effect can be achieved using either the F\"orster resonance or vdw interaction, as confirmed by experimental observations. \cite{PhysRevLett.97.083003,urban2009observation,gaetan2009observation,PhysRevLett.119.160502} However, in a many-body system, the construction of a Rydberg superatom is typically facilitated by the Rydberg interaction in the form of a pairwise additive vdw potential.\cite{PhysRevLett.99.163601,PhysRevLett.100.033601,dudin2012observation,PhysRevX.5.031015,PhysRevResearch.2.043339} This is particularly true for the Rydberg-$S$ state, which possesses only one repulsive branch in its molecular potential, preventing a breakdown of the Rydberg blockade.\cite{Walker_2005,PhysRevLett.102.013004,PhysRevA.91.023411,PhysRevA.92.063419,PhysRevLett.120.113602}
Here, we will provide a concise theoretical analysis of the mechanism and associated properties of Rydberg superatoms. Detailed explanations of the experimental procedures and their applications in quantum information and quantum optics will be discussed in the following sections.

\begin{figure}
\centering  
\includegraphics[width=0.9\linewidth]{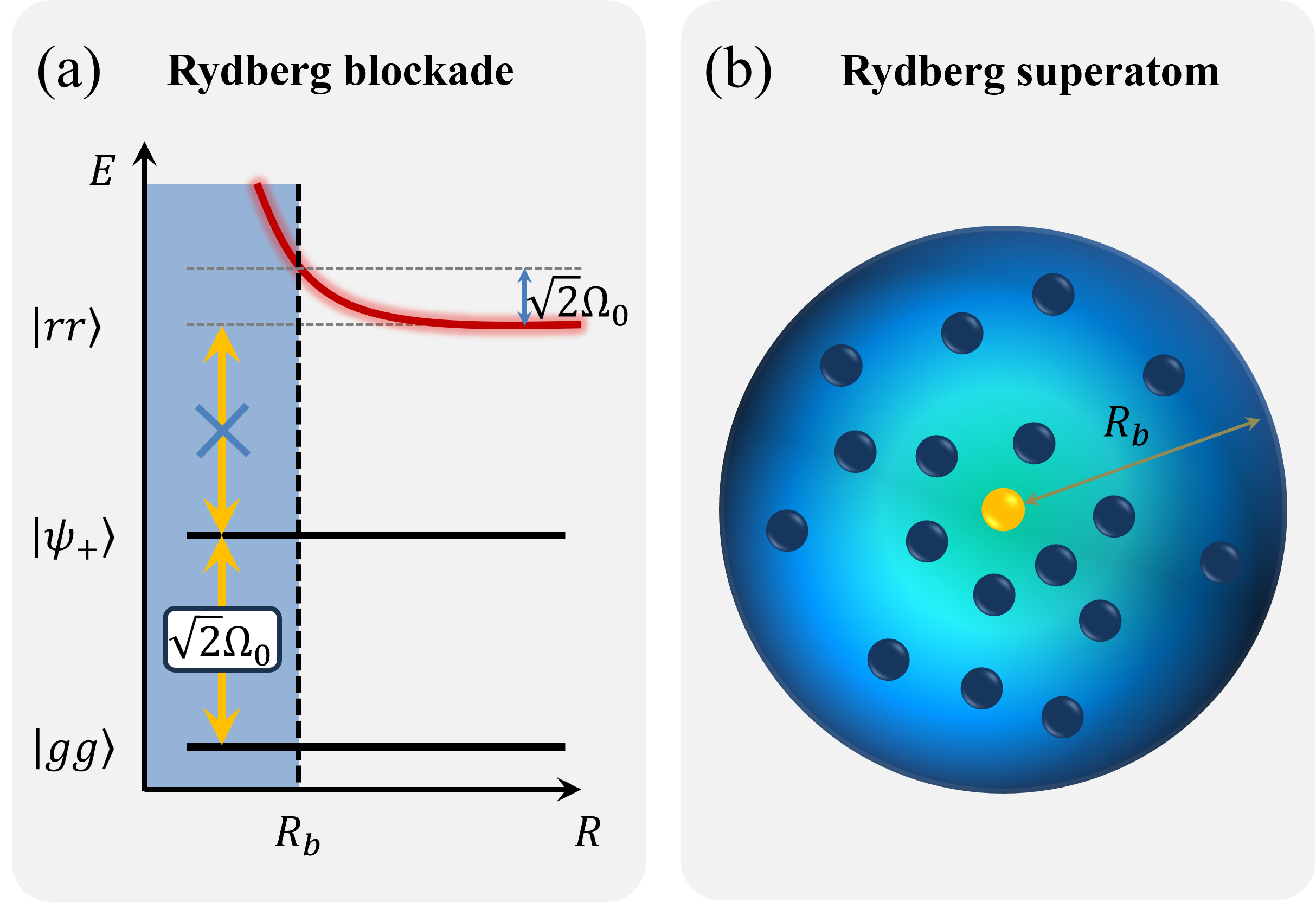}
\caption{\label{psuperatom}(a) Rydberg blockade mechanism. The double excitation of the Rydberg state $|rr\rangle$ is inhibited within the blockade radius $R_b$ owing to the large energy shift induced by the strong repulsive interaction. (b) Schematic view of a Rydberg superatom comprised of $N$ atoms sharing a single Rydberg excitation enclosed within a blockade sphere. 
}
\end{figure}

Consider an ensemble of $N$ two-level atoms, with ground state $|g\rangle$ and Rydberg state $|r\rangle$, being illuminated by a resonant laser field characterized by a Rabi frequency $\Omega_0$. In the interaction picture, the Hamiltonian of the system reads
\begin{equation}\label{superatom}
\hat{H}_{I}=\sum_{i=1}^N\frac{\hbar\Omega_0}{2}|g_i\rangle\langle r_i|+{\rm H.c.}-\frac{1}{2}\sum_{i\neq j}\frac{C_6}{R_{ij}^6}|r_ir_j\rangle\langle r_ir_j|,
\end{equation}
where a negative value of $C_6$ denotes repulsive vdw interactions, while a positive value signifies attractive interactions. At the same time, for the sake of simplification, we temporarily disregard the impact of atomic position and motion on the relative phase between individuals. Under the influence of Eq.~(\ref{superatom}), if all atoms in the ensemble are initially set in the ground state $|G\rangle=|g_1,g_2,...,g_N\rangle$, they will collectively transition to a symmetric single atom excitation state, assuming $C_6/R_{ij}^6\gg\hbar\Omega_0$. 
{When examining the simple case of the collective excitation of two atoms, as illustrated in Fig.~\ref{psuperatom}(a), the associated collective blockade radius is defined as $R_b=[C_6/(\hbar\sqrt{2}\Omega_0)]^{1/6}$,\cite{PhysRevLett.107.093601} 
where the interaction energy equals the
linewidth of excitation dominated
by the power broadening $\sqrt{2}\Omega_0$.}
Within this range, Rydberg atoms manifest a robust repulsive interaction, constraining excitation to a single excited symmetric state $|\psi_+\rangle=(|gr\rangle+|rg\rangle)/\sqrt{2}$ for the pair of atoms,  assuming both atoms are driven with the same Rabi frequency. An offset in Rabi frequencies of two atoms may freeze the dynamics in a single atom, inhibiting the creation of the collective single excited state.\cite{Srivastava_2019}
Expanding the system to the collective excitation of $N$ atoms, the collective blockade radius should be calculated as 
$R_b=[C_6/(\hbar\sqrt{N}\Omega_0)]^{1/6}$, and the corresponding collective state is represented by
\begin{equation}\label{superatomstate}
|R\rangle=\frac{1}{\sqrt{N}}\sum_{j=1}^N|g_1,g_2,...,r_j,...,g_N\rangle,
\end{equation}
where all atoms share a single Rydberg excitation. {This form of collective excitation also exists in the single-photon superradiance model, where identical ground state atoms interact with a single photon.\cite{PhysRevLett.96.010501,PhysRevLett.100.160504,PhysRevLett.102.143601,PhysRevLett.115.243602,PhysRevLett.114.043602}}
Figure~\ref{psuperatom}(b) presents a schematic representation of a Rydberg superatom, which is collectively coupled to the ground state $|G\rangle$ with a coupling strength of $\sqrt{N}\Omega_0$.
We note that the Rydberg superatom is itself a multiparticle $W$ state with maximum entanglement.\cite{PhysRevA.62.062314} This entangled state holds significant relevance in quantum information processing.\cite{PhysRevA.65.032108,RevModPhys.81.865,Kim2022} Recent experiments has demonstrated its robustness to both particle loss and external electric field noise, particularly as a logical qubit.\cite{PhysRevLett.127.050501,PhysRevLett.127.063604}
Therefore, the emergent properties of a Rydberg superatom are derived from the collective behavior of its constituent atoms rather than from the properties of individual atoms alone. These properties encompass enhanced nonlinear optical responses and strong light-matter coupling, both of which hold promise for applications in quantum information processing and quantum optics. As the Rydberg superatom system progresses to the forefront of quantum science, it demonstrates the potential to revolutionize various technological domains, including quantum computing and advanced materials engineering. {One important theme in the ongoing research is to develop experimental techniques and explore theoretical models to understand and manipulate properties and dynamics of Rydberg superatoms.}

In the context of the Rydberg atom system, along with the Rydberg blockade, phenomena such as the Rydberg antiblockade,\cite{PhysRevLett.98.023002,PhysRevLett.104.013001,Bai_2020} Rydberg dressing,\cite{PhysRevLett.104.195302,PhysRevLett.104.223002,PhysRevLett.105.160404,PhysRevA.82.033412,PhysRevA.85.053615,Balewski_2014,jau2016entangling} ground state blockade,\cite{PhysRevA.96.012328,PhysRevLett.125.073601} and unconventional Rydberg pumping are encountered.\cite{PhysRevA.98.062338,PhysRevA.102.053118} While these approaches diversify how Rydberg atoms interact, they fall outside the purview of Rydberg superatoms and, therefore, lie beyond the scope of our discussion here.

\section{EXPERIMENTAL TECHNIQUES AND ADVANCES}
\label{sec:experimentaltech}

\subsection{Coherent laser excitation of Rydberg atoms}

An essential step in Rydberg experiments is the preparation of Rydberg atoms. The utilization of lasers offers high-resolution, coherent, and efficient excitation of Rydberg states. 
After the invention of tunable dye lasers, Rydberg laser spectroscopy with resolution $\sim 0.1\rm~nm$ was demonstrated in 1973.\cite{J.Phys.B.atom.molec.phys.vol6.august.1973, 1983.in.Rydberg.States.of.Atoms.and.Molecules}
The spectral resolution was improved to $\sim$ 10~GHz in 1979.\cite{PhysRevA.20.2251}
In 2003, Rydberg Autler-Townes spectra with~MHz resolution were studied with narrow linewidth lasers. \cite{PhysRevA.68.053407}

\begin{figure}
\centering  
\includegraphics[width=1\linewidth]{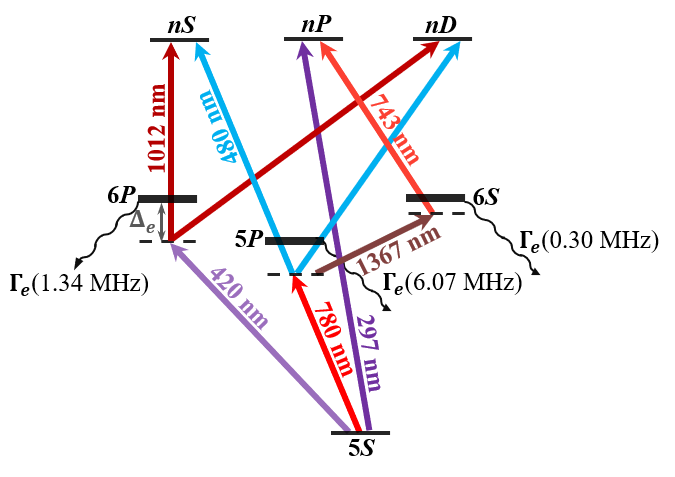}
\caption{\label{LLfig3}Different Rydberg laser excitation schemes for $^{87}$Rb atom. 297~nm UV laser can couple the ground state $5S$ and the Rydberg state $nP$. 780~nm (420~nm) and 480~nm (1012~nm) lasers can excite atoms from $5S$ to $nS$ or $nD$ via an intermediate state $5P$ $(6P)$ with intermediate detuning $\Delta_{e}$. Due to the finite lifetime of the intermediate states, atoms in these states decay with a rate of $\Gamma_{e}$, which is one of the main sources of error in the two-photon excitation scheme. Using the intermediate states $5P$ and $6S$, atoms can be excited to the $nP$ Rydberg states using the 780~nm, 1367~nm, and 743~nm lasers.}
\end{figure}

\begin{figure*}
\centering  
\includegraphics[width=0.7\linewidth]{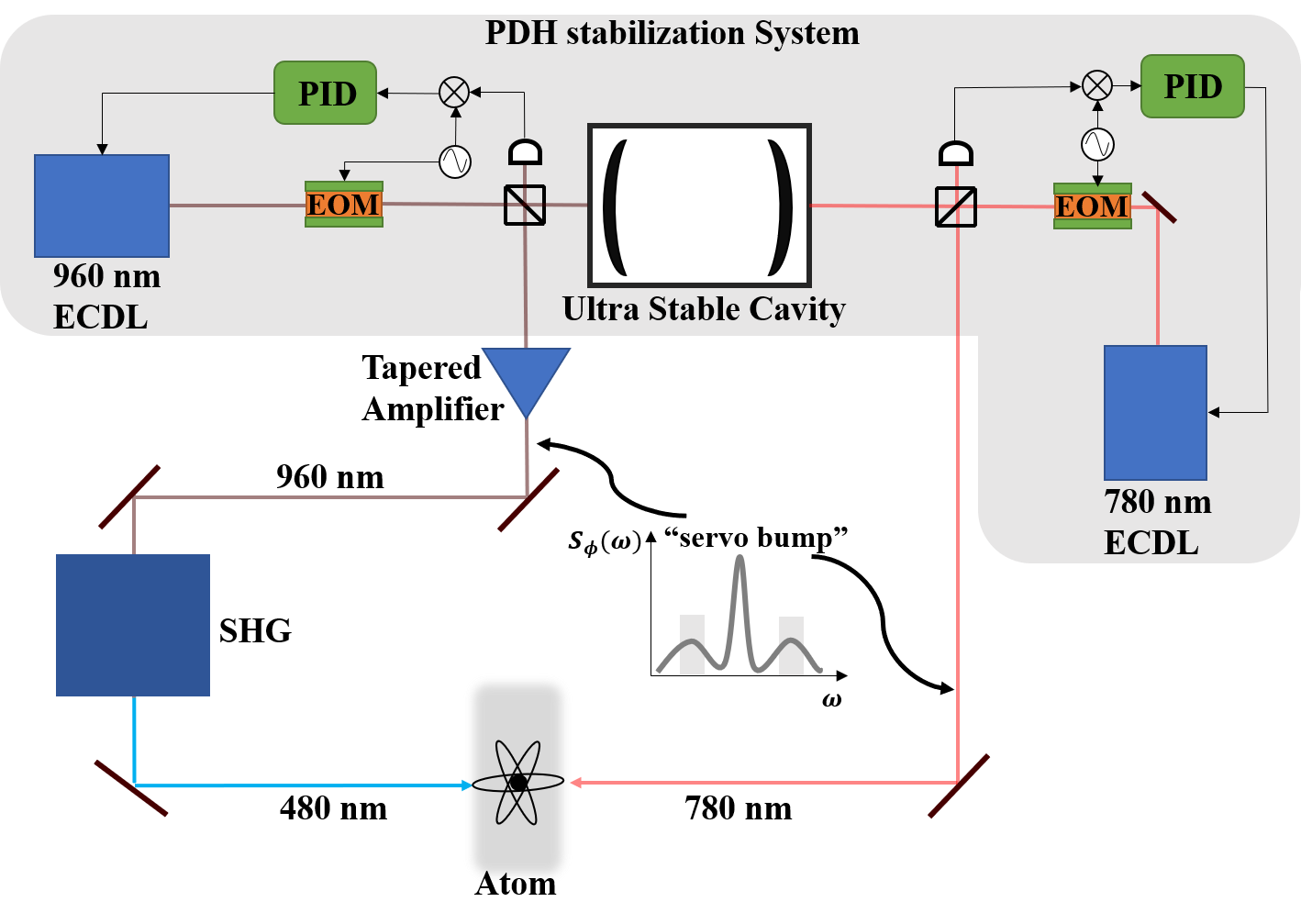}
\caption{\label{LLfig4} Laser setup of 480~nm and 780~nm lasers used in $^{87}$Rb Rydberg excitation. The 780~nm and 960~nm lasers are obtained from external cavity diode lasers (ECDLs). Both lasers are frequency locked to a high finesse, ultra-stable cavity using the Pound-Drever-Hall (PDH) locking technique, which generates ``servo bump'' due to the limited stabilization bandwidth. Frequency doubling of the power-amplified (using the tapered amplifier) 960~nm laser in the second harmonic generator (SHG) generates blue light at 480~nm. The 780~nm and 480~nm laser beams excite the atoms in a counterpropagating configuration to minimize the Doppler effect.}
\end{figure*}

The laser fields used for Rydberg excitation must be intense enough to guarantee a fast excitation within the Rydberg lifetime and coherence time. The natural lifetimes of high-lying Rydberg states are mainly limited by spontaneous emission and black-body radiation, and can last more than hundreds of microseconds. However, the ground-Rydberg coherence times are usually about a few to tens of microseconds, limited by many decoherence channels, such as the Doppler effect, density-dependent phase shift, fluctuating background electric fields, etc. Therefore, the Rabi frequencies of the excitation laser have to be large, usually on the orders of~MHz, to ensure coherent excitations. Such large Rabi frequencies require intense light sources for highly excited Rydberg states since the dipole matrix elements between the ground and Rydberg states decrease with the principal quantum number as $n^{*-1.5}$.
In Rydberg tweezer array experiments, it has been shown that a larger Rabi frequency is crucial for high-fidelity operations within the finite ground-Rydberg coherence.\cite{Madjarov2020} In a Rydberg superatom containing $N$ atoms, the effective ground-Rydberg Rabi frequency is improved by a factor of $\sqrt{N}$ due to collective enhancement effects.

Given the considerable energy gaps between the ground and Rydberg states, direct ground-to-Rydberg transition frequencies typically fall within the ultra-violet (UV) range.\cite{Wang:17}
Taking Rubidium atoms as an example, lasers with $\sim$ 297~nm wavelength can {excite} atoms from the $5S$ ground state to the $nP$ Rydberg states.\cite{Thoumany:09} Coherent manipulation between the ground and Rydberg states using a UV laser was demonstrated in Fig.~3, see Ref.~\onlinecite{PhysRevA.89.033416}. The single photon UV scheme is ideal for realizing the Rydberg dressing and has also been used in the ground-Rydberg state quantum memory.\cite{Li2016}

{The choice of the excitation scheme depends on the experimental aim.
A commonly adopted scheme involves an intermediate state and uses two-photon Rydberg excitation, which can benefit from electromagnetically induced transparency (EIT).  With the 780~nm-480~nm (420~nm-1012~nm) two-photon scheme, the ground state Rb atom can be excited to the $nS$ or $nD$ Rydberg states, as shown in Fig.~\ref{LLfig3}. }
Figure~\ref{LLfig4} shows the laser setup for the 780~nm-480~nm scheme. The 780~nm light field can be easily obtained from a commercially available external cavity diode laser (ECDL), while the 480~nm light comes from a second-harmonic-generator (SHG) seeded by power-amplified 960~nm laser light.
To achieve a strong blockade effect, the laser linewidth must be narrow enough compared to the Rydberg interactions. Therefore, both 780~nm and 960~nm ECDLs are frequency stabilized to an ultralow expansion cavity with high finesse to guarantee narrow linewidth. However, the Pound-Drever-Hall (PDH) laser locking technique leads to large ``servo bumps'' outside the stabilization bandwidth, which could cause severe decoherence during laser excitation. 

Besides the above-mentioned single and two-photon schemes, three-photon Rydberg laser excitation is also possible. A technical advantage of the three-photon protocol is that it is possible to choose certain intermediate states, such that all of the lasers involved in the excitation are in the near-infrared (NIR) regime.
For $^{87}\rm Rb$, as shown in Fig.~\ref{LLfig3}, lasers at 780~nm, 1367~nm, and 743~nm can be used for Rydberg excitation through two intermediate states 5$P$ and 6$S$. Furthermore, by carefully arranging the direction of the excitation lasers, Doppler-free Rydberg excitation can be achieved using the three-photon scheme.\cite{PhysRevA.84.053409}

\begin{figure}
\centering  
\includegraphics[width=1\linewidth]{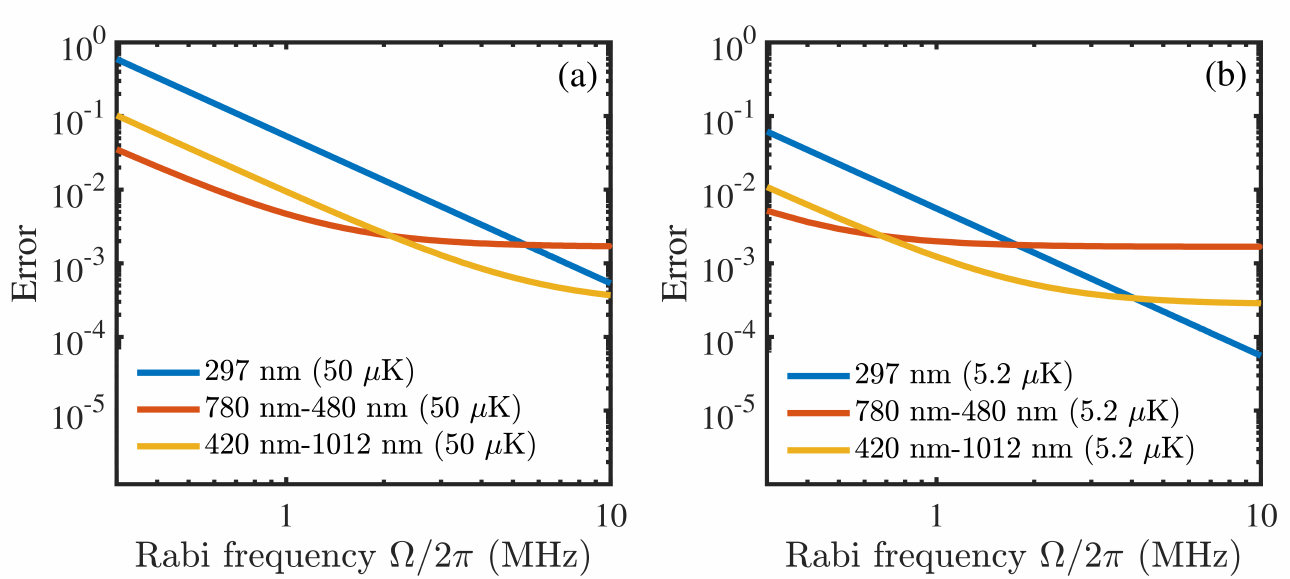}
\caption{\label{LLfig5} Influence of Rabi frequency ($\Omega_{1}=\Omega_{2}$ for two-photon excitation schemes) on { Doppler-induced} Rydberg excitation error in different schemes for an atomic temperature of (a) ${50}~{\mu}$K and (b) ${5.2}~{\mu}$K.  The simulation is performed with intermediate detuning $\Delta_{e}$ = $2\pi\times5687$~MHz in 780~nm-480~nm scheme and $2\pi\times7.6$~MHz in 420~nm-1012~nm scheme.}
\end{figure}

The Rydberg excitation fidelity is affected by various decoherence and loss mechanisms.
For the single-photon scheme, a major source of errors in coherent excitation between ground and Rydberg states is the Doppler effect due to atomic motions, which results in a random detuning $\delta=k v$, where $v$ is the velocity of the atom along the wave vector of the laser beam $k$. Limited by the available UV laser power and the small dipole moment between the ground and Rydberg states, the Rabi frequency of the single-photon excitation {$\Omega_{\rm UV}$ is on the order of a few~MHz. As shown in Fig.~\ref{LLfig5}(a), the ground-Rydberg states $\pi$-pulse excitation error exceeds a few percent for an atomic temperature of ${50}~{\mu}$K and $\Omega_{\rm UV}\sim$~MHz}. Infidelity caused by the Doppler effect can be suppressed by lowering the temperature or increasing the Rabi frequency.

Compared to the UV laser in single-photon excitation, it is easier to achieve high power for the visible and near-infrared lasers used in two-photon schemes. Therefore, the Rabi frequency of the ground-Rydberg state $\Omega=\Omega_{1}\Omega_{2}/(2\Delta_{e})$ in two-photon excitation can reach tens or even hundreds of~MHz, where $\Omega_{1}$ $(\Omega_{2})$ is the Rabi frequency for the ground-intermediate (intermediate-Rydberg) transition, and $\Delta_{e}$ is the intermediate-state detuning, as shown in Fig.~\ref{LLfig3}.

In the two-photon scheme, if a counterpropagating excitation protocol is employed, the Doppler effect will be less severe than in the single-photon case.
As shown in Fig.~\ref{LLfig4}, {by using counterpropagating excitation beams, the Doppler effect can be minimized.\cite{PhysRevLett.121.123603}}
It is also obvious that the 780~nm-480~nm excitation scheme has a smaller Doppler-induced error than that of the 420~nm-1012~nm scheme. However, regardless of the excitation approach used, achieving a Doppler-induced error well below 1\% is achievable by either lowering the atomic temperature or applying a strong laser (large Rabi frequency).

Due to the finite lifetime of intermediate states, the Rydberg excitation fidelity is also affected by the spontaneous emission.\cite{Browaeys_2016} The excitation error for a ground-Rydberg $\pi$ pulse is $\sim$ $ \frac{\pi\Gamma_{e}}{4\Delta_{e}}(\frac{\Omega_{1}}{\Omega_{2}}+\frac{\Omega_{2}}{\Omega_{1}})$, where $\Gamma_{e}$ is the spontaneous emission rate of the intermediate state. For a fixed detuning $\Delta_{e}$, this error is minimized when $\Omega_{1}=\Omega_{2}$. 
Since the dipole matrix element between the ground and intermediate states is orders of magnitude larger than that between the intermediate and Rydberg states, the two-photon Rabi frequency is limited by the achievable laser power for the intermediate-Rydberg transition.
Recent developments in fiber lasers make it possible to achieve tens or even hundreds of watts of power at 1012~nm.\cite{PhysRevA.98.033411}
Meanwhile, 6$P_{3/2}$ has a lower spontaneous emission rate than 5$P_{3/2}$.  Consequently, the error caused by the spontaneous emission of the intermediate state in the 420~nm-1012~nm scheme can be orders of magnitude lower than that in the 780~nm-480~nm scheme.

\begin{figure}
\centering  
\includegraphics[width=0.95\linewidth]{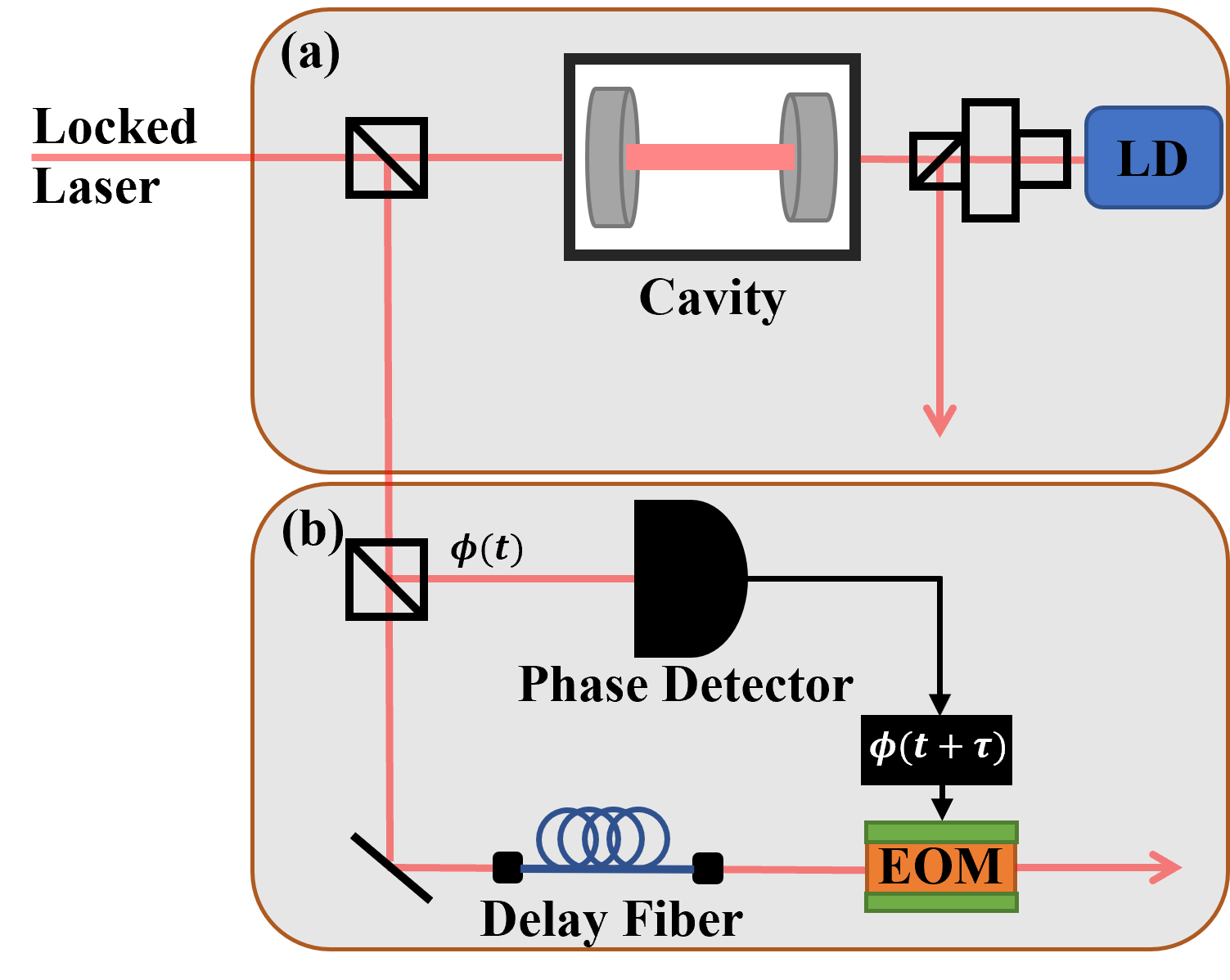}
\caption{\label{LLfig6} Filtering Cavity and feedforward techniques for suppressing the ``servo bump''. (a) The frequency-stabilized laser is sent to a narrow bandwidth optical cavity. The ``servo bump'' phase noise in the transmitted laser will be strongly suppressed, which is used for the injection lock of another laser diode (LD). (b) The phase information of the laser is obtained by a phase detector, which is then feedforward to an electro-optic modulator (EOM) to cancel the phase noise of the laser. The delay fiber is used to minimize the inherent delay time $\tau$ of the phase detector.}
\end{figure}

As illustrated in Fig.~\ref{LLfig5}, at  $50~{\mu}$K and $5.2~{\mu}$K, we conduct a comparative analysis of the errors using three different laser excitation schemes with varying Rabi frequencies. For the 780~nm-480~nm excitation scheme, the intermediate state detuning is set at $2\pi\times5687$~MHz,\cite{PhysRevA.105.042430} while for the 420~nm-1012~nm,  it is $2\pi\times7.6$ GHz.~\cite{Evered2023} From the simulation, it is evident that Doppler-induced infidelity can be effectively suppressed by increasing the excitation Rabi frequency.

In addition, laser phase noise and intensity noise also contribute to the Rydberg excitation error. Depending on the laser phase stability and the Rabi frequency, the contribution of phase noise to the excitation error could be more than a few percent~\cite{PhysRevA.97.053803} without suppressing the servo bump. By using an optical cavity as a filter, as shown in Fig.~\ref{LLfig6}(a), the servo bump, which is typically around~MHz, is strongly suppressed by the cavity since its transmission bandwidth can be narrower than kHz.~\cite{PhysRevLett.121.123603} Moreover, Fig.~\ref{LLfig6}(b) shows an alternative protocol, the feedforward technique, for suppressing the servo bump. A phase discrimination circuit based on the beat signal of the laser~\cite{PhysRevApplied.18.064005} and the residual PDH error signal is used as the phase detector. \cite{chao2023pounddreverhall} The signal obtained from the phase detector is then fed forward to an electro-optical modulator (EOM) with optimized modulation depth so the phase noise of the laser can be reduced. An appropriate length of delay fiber is used to overcome the frequency deviations caused by the inherent delay time of the phase detector. It is reported that after suppressing the servo bump, the fidelity after $9\pi$ pulse ~\cite{PhysRevApplied.18.064005}  and the coherence time~\cite{PhysRevLett.121.123603} can be significantly increased. 
The intensity noise largely depends on the lasers themselves. For example, the 297~nm laser used in the single-photon scheme in $^{87}$Rb atoms can be obtained from the fourth harmonic generation seeded {by a power-amplified 1188~nm laser.\cite{PhysRevA.99.042502}} Using stimulated Raman scattering fiber amplifier technology, the amplified 1188~nm laser usually has a peak-to-peak intensity noise of a few percent. In addition, the cavities for frequency quadrupling will introduce additional intensity noise in the cavity locking. Considering the above noise sources, the resulting Rydberg excitation errors could be as large as a few percent.

\subsection{Preparation of Rydberg superatoms: Many-body Rabi oscillation}

{The concept of the Rydberg superatom can be connected to the original idea proposed by Dicke, where 
$N$ atoms are coupled to a single mode of the electromagnetic field at small interatomic distances. In the Dicke model, the collective behavior of closely spaced atoms leads to phenomenon such as superradiance, where the atoms act coherently as a single quantum system.\cite{dicke1954coherence} This concept can be extended to the Rydberg superatom, where the interatomic distances can be much larger than the wavelength of the excitation light. Despite the larger distances, the atoms in a Rydberg superatom can still form collective excitation states, allowing them to behave as a single, coherent entity.} 
Such a collective effect could lead to a many-body Rabi oscillation with an enhanced Rabi frequency of $\sqrt{N}$$\Omega$.\cite{rabi1939molecular,cummings1983exact,Kumlin_2023}
In 2001, Lukin {\it et al.} suggested that if an atomic ensemble is confined within the blockade radius,\cite{PhysRevLett.87.037901} many-body Rabi oscillation could be achieved by performing laser excitation between the ground state $\ket{G}$ and the Rydberg superatom state $\ket{R}$ of Eq.~(\ref{superatomstate}). Coherent many-body Rabi oscillation is not only conclusive evidence for the successful preparation of the Rydberg superatom but also the cornerstone for many Rydberg-atom-related quantum applications, such as deterministic quantum photonic state preparation and manipulation, quantum networks, and quantum simulations.

\begin{figure}
\centering  
\includegraphics[width=0.9\linewidth]{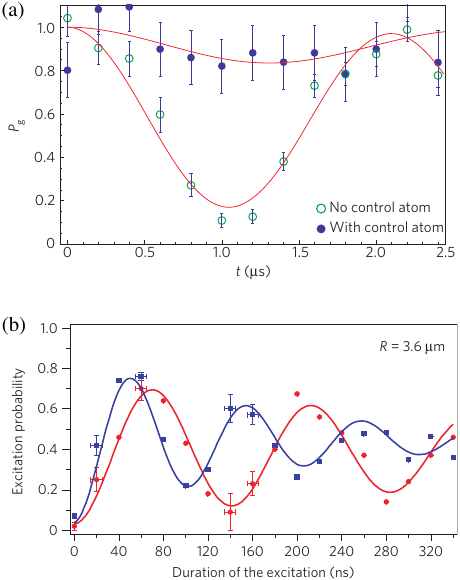}
\caption{\label{LLfig7}(a) Rydberg blockade between two single atoms.
  Reproduced with permission from Urban
{\it et al.}, Nature Physics {\bf 5}(2), 110-114 (2009). Copyright 2009 Springer Nature.\cite{urban2009observation} (b) Rabi oscillations enhanced by two atoms. Reproduced with permission from Ga\"etan {\it et al.}, Nature Physics {\bf 5}(2), 115-118 (2009). Copyright 2009 Springer Nature.\cite{gaetan2009observation}}
\end{figure}

Although Rydberg interactions were already observed in the 1980s, Rydberg excitation blockade between two single atoms was not demonstrated until 2009.\cite{urban2009observation, gaetan2009observation} As illustrated in Fig.~\ref{LLfig7}(a), the unfilled circles show the ground-Rydberg state Rabi oscillation of a single atom confined in $\mu$m sized far-off-resonance trap.
When a control atom, which is $\sim$ 10 $\mu \rm m$  apart, is driven to the Rydberg state, the excitation of the target atom (filled circles) is suppressed due to a strong two-atom interaction shift of about 9.5~MHz. \cite{urban2009observation}
This conditioned Rydberg excitation blockade mechanism paves the way for the two-qubit quantum logic gate with neutral atoms.\cite{PhysRevLett.104.010503}

Figure~\ref{LLfig7}(b) demonstrates the first two-atom enhanced ground-Rydberg Rabi oscillation.\cite{gaetan2009observation}
To achieve a strong Rydberg blockade, two atoms are brought within a distance of 4 $\mu \rm m$. Careful selection of Rydberg levels results in the F\"orster resonance, leading to a Rydberg-Rydberg interaction strength $\sim$ 50~MHz. 
The Rabi oscillation of two atoms (blue squares) is accelerated compared to the case of a single atom (red circles), by a factor of 1.38(3), close to the expected value of $\sqrt{2}$ in the two-atom blockade model.

\begin{figure*}
\centering  
\includegraphics[width=0.8\linewidth]{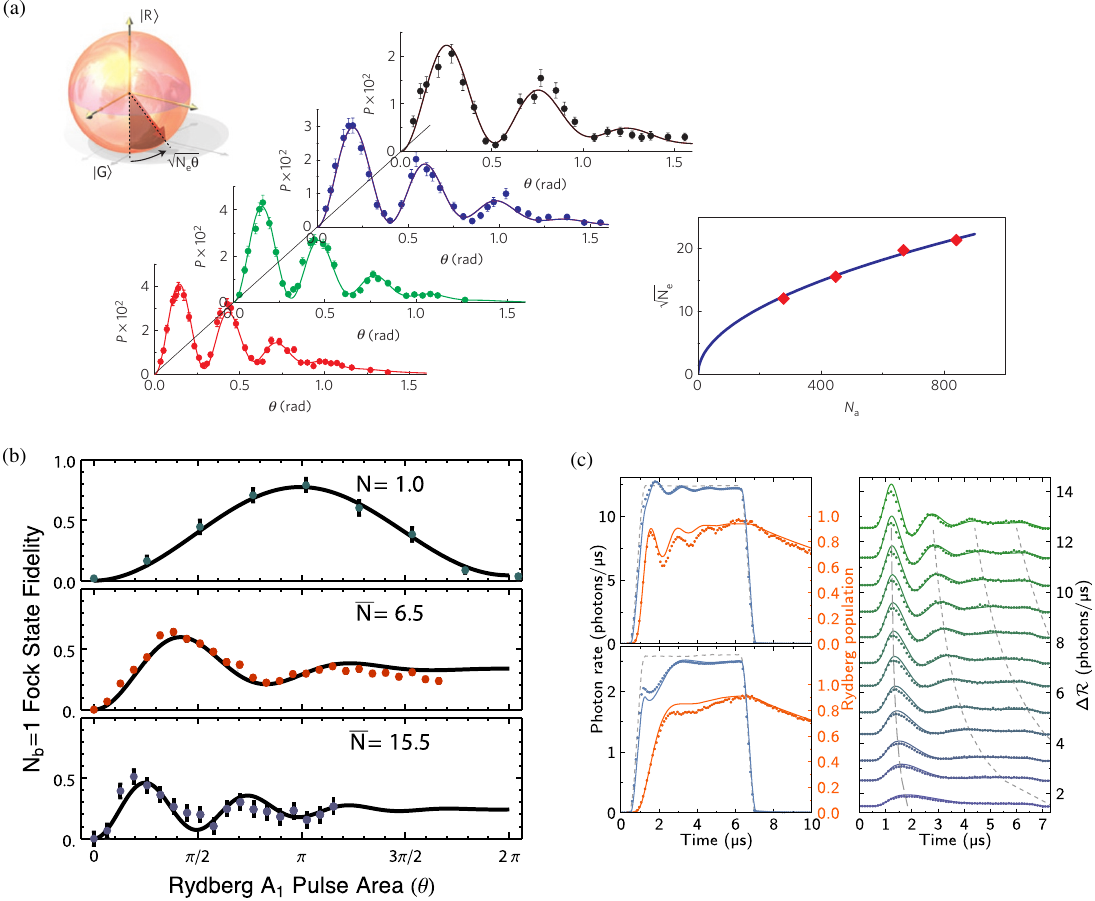}
\caption{\label{LLfig8}Many-body Rabi oscillations in atomic ensembles.
(a) Coherent many-body Rabi oscillations of a mesoscopic atomic ensemble. Reproduced with permission from Dudin
{\it et al.}, Nature Physics {\bf 8}(11), 790-794 (2012). Copyright 2012 Springer Nature.\cite{dudin2012observation} 
(b) Rabi oscillation between the ground state and superatom state for various atom number distributions. Reproduced with permission from
Ebert {\it et al.}, Phys. Rev. Lett. {\bf 112}(4), 043602 (2014). Copyright 2014
American Physical Society.\cite{PhysRevLett.112.043602}
(c) Time evolution of the photon signal and the Rydberg population.  Reproduced with permission from 
Paris-Mandoki {\it et al.}, Phys. Rev. X {\bf 7}(4), 041010 (2017). Copyright 2017
American Physical Society, licensed under a Creative Commons Attribution 4.0 International license.\cite{PhysRevX.7.041010}}
\end{figure*}

In 2012, a coherent many-body Rabi oscillation was demonstrated in a mesoscopic ensemble containing a few hundred atoms.\cite{dudin2012observation}
To achieve blockade over the entire ensemble, the atoms are confined within a region $\sim$ 6 $\mu \rm m$ and a high-lying Rydberg state $\ket{102S_{1/2}}$ is used.
As shown in Fig.~\ref{LLfig8}(a), pronounced many-body Rabi oscillations are observed between the ground state $\ket{G}$ and the Rydberg superatom state $\ket{R}$  under different effective atom numbers $N_e$.
From the oscillation dynamics, the collective enhancement factor can be extracted and compared with the atom number $N_a$.
The observed $\sqrt{N_e}$ enhancement behavior agrees with Dicke's prediction. 
The detection of the Rydberg state is accomplished by converting the superatom state into a single photon. Therefore, this work provides the basis for high-quality Rydberg single-photon source and related quantum optics research.

Several experiments conducted on atomic ensembles have also confirmed a similar many-body Rabi oscillation phenomenon. 
Figure~\ref{LLfig8}(b) shows the many-body Rabi oscillations where the number of atoms ranges from 1 to 15.\cite{PhysRevLett.112.043602} After the preparation of the Rydberg superatom $\ket{R}$, it can be transferred to a ground state superatom $\ket{G'}$. 
The above protocol can be repeated to deterministically prepare the atomic Fock state $\ket{G'}^n$.
Figure~\ref{LLfig8}(c) displays the many-body Rabi oscillation under strong dissipation,\cite{PhysRevX.7.041010} which facilitates the realization of a Rydberg-based single-photon absorber.\cite{PhysRevLett.117.223001}

\begin{figure*}
\centering  
\includegraphics[width=0.8\linewidth]{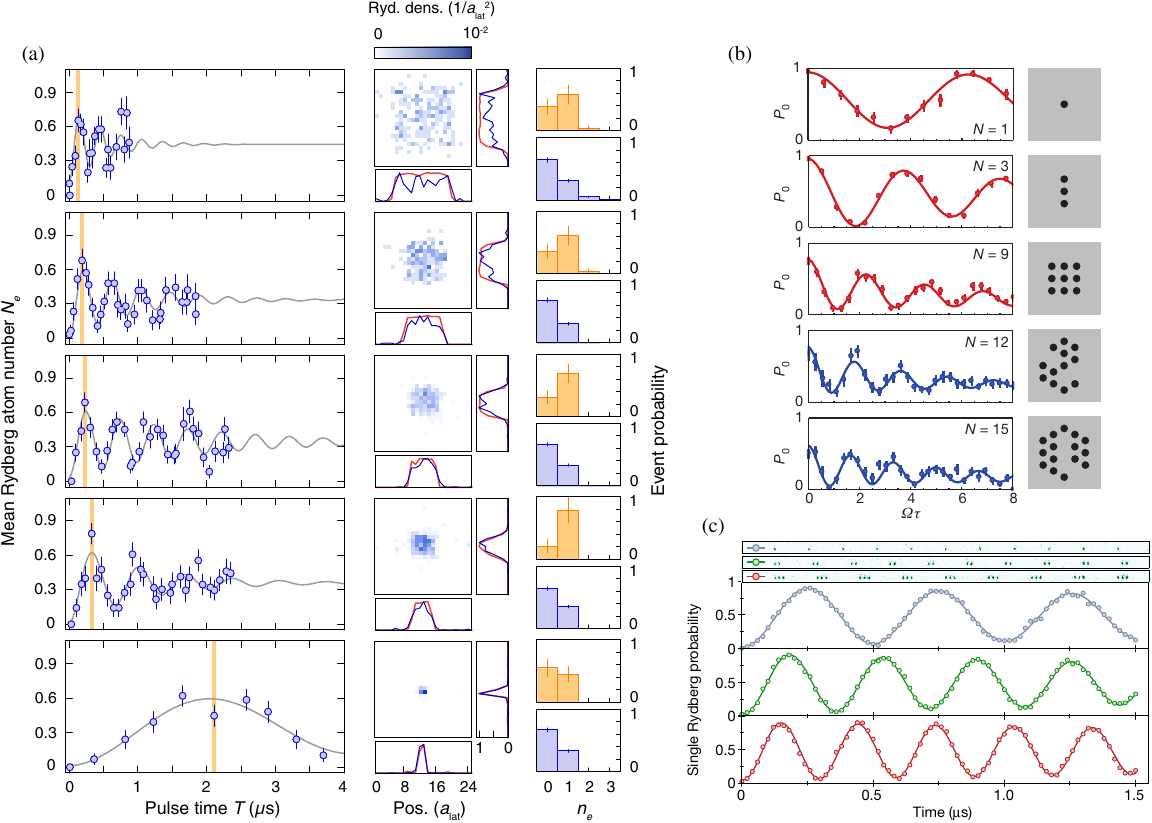}
\caption{\label{LLfig9}Many-body Rabi oscillations in atom array experiments. 
(a) Collective Rabi oscillations in a quantum gas microscope. Reproduced with permission from 
Zeiher {\it et al.}, Phys. Rev. X {\bf 5}(3), 031015 (2015). Copyright 2015
American Physical Society, licensed under a Creative Commons Attribution 3.0 License.\cite{PhysRevX.5.031015}
(b) Collective Rabi oscillation in two-dimensional atomic arrays. Reproduced with permission from  Labuhn
{\it et al.}, Nature {\bf 534}(7609), 667-670 (2016). Copyright 2012 Springer Nature.\cite{labuhn2016tunable}
(c) Collective Rabi oscillation in one-dimensional atomic arrays. Reproduced with permission from  Bernien
{\it et al.}, Nature {\bf 551}(7682), 579-584 (2017). Copyright 2012 Springer Nature.\cite{bernien2017probing}}
\end{figure*}

In addition to atomic ensembles, many-body Rabi oscillations have also been observed in atom array settings. In 2015, Zeiher {\it et al.} demonstrated the preparation of a scalable Rydberg superatom in a quantum gas microscope [Fig.~\ref{LLfig9}(a)].\cite{PhysRevX.5.031015}
The $\sqrt{N}$ scaling in many-body Rabi oscillation was observed from one to a few hundred atoms.
Related advances have also been made in reconfigurable tweezer arrays. 
Figures~\ref{LLfig9}(b) and \ref{LLfig9}(c) show the many-body Rabi oscillation with high contrast in one- and two-dimensional atomic arrays.\cite{labuhn2016tunable, bernien2017probing}

{\subsection{Quantum interface and coherence property}}

\begin{figure}
  \centering  
  \includegraphics[width=1\linewidth]{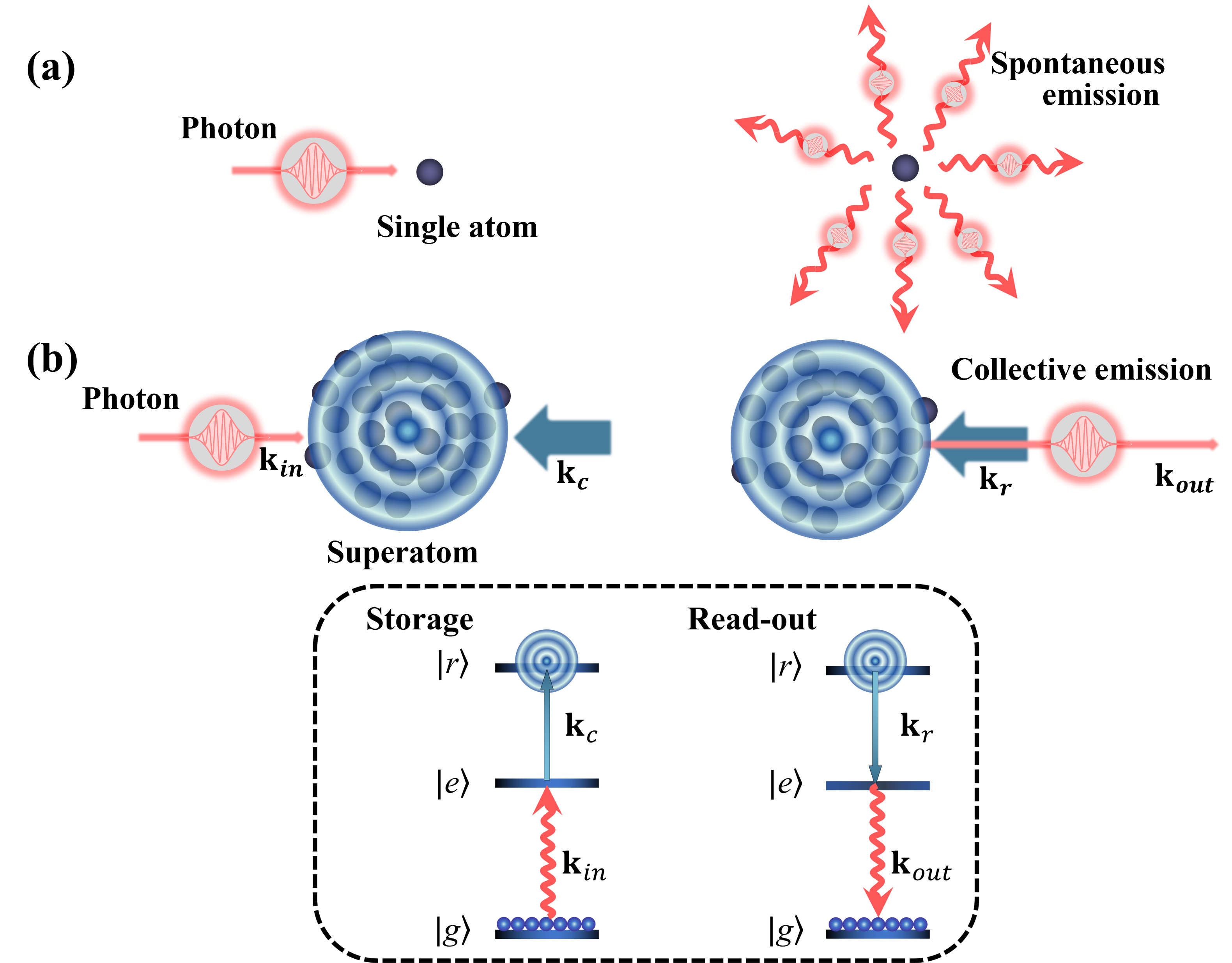}
  \caption{
  Efficient matter-light quantum interface using Rydberg superatom. (a) A single atom adsorbs and re-emits a single photon in free space. (b) Storage and retrieval of a single photon using Rydberg superatom. The inset shows the relevant atomic levels.
  }
  \label{LLfig10}
\end{figure}

One of the most prominent features of Rydberg superatoms is their excellent matter-light quantum interface,\cite{ PhysRevLett.87.037901, PhysRevA.66.065403, vuletic2006superatoms} as shown in Fig.~\ref{LLfig10}. When a single atom absorbs and re-emits a single photon in free space, it is dominated by the spontaneous emission process, and the photon is ejected into 4$\pi$ solid angles. In stark contrast, a Rydberg superatom can emit single photons into a highly directional spatial mode as a result of the collective enhancement. Figure~\ref{LLfig10}(b) illustrates the single photon storage-and-retrieval protocol using a Rydberg superatom. 
Using a control laser connecting the intermediate state $\left|e\right\rangle$ and a Rydberg state $\left|r\right\rangle$, Rydberg EIT storage can be performed, and a single photon resonant with the $\left|g\right\rangle\leftrightarrow\left|e\right\rangle$ transition can be coherently transferred to a Rydberg superatom, leaving the field in the vacuum state\cite{PhysRevLett.96.010501}
\begin{equation}
\label{2-photons-Rydberg}
\frac{1}{\sqrt{N}} \sum_{j=1}^N e^{i\left(\mathbf{k}_{in}+\mathbf{k}_{c}\right) \cdot \mathbf{x}_j}|g_1,g_2,...,r_j,...,g_N\rangle\otimes|0\rangle.
\end{equation}
Here, $\mathbf{x}_j$ is the position of the $j$-th atom, and $\mathbf{k}_{in}, \mathbf{k}_{c}$ are the wave vectors of the input photon and the control laser, respectively. $|0\rangle$ is the vacuum state of the photon field.
This Rydberg EIT light storage process efficiently converts a photon into a single Rydberg excitation collectively shared among the atoms inside the Rydberg blockade radius.

Within the coherence time of the Rydberg superatom, the stored excitation can be transferred back to a single photon, with a read-out laser field. The read-out laser transfers the collective excitation from the Rydberg superatom state to the intermediate state,
\begin{equation}
\label{2-photons-e}
\frac{1}{\sqrt{N}} \sum_{j=1}^N e^{i\left(\mathbf{k}_{in}+\mathbf{k}_{c}-\mathbf{k}_{r}\right) \cdot \mathbf{x}_j}|g_1,g_2,...,e_j,...,g_N\rangle \otimes|0\rangle,
\end{equation}
where $\mathbf{k}_{r}$ is the wave vector of the readout laser.
Due to the finite lifetime in the intermediate state $\left|e\right\rangle$, the collective state  soon radiates into a new state,
\begin{equation}
\label{2-photons-g}
\frac{1}{\sqrt{N}} \sum_{j=1}^N e^{i\left(\mathbf{k}_{in}+\mathbf{k}_{c}-\mathbf{k}_{r}-\mathbf{k}_{out}\right) \cdot \mathbf{x}_j}|g_1,g_2,...,g_j,...,g_N\rangle\otimes|1\rangle_{\mathbf{k}_{out}},
\end{equation}
where the atoms are back in the ground state $\left|g\right\rangle$ and a single photon is emitted into the direction defined by its wave vector $\mathbf{k}_{out}$. 
The probability of emitting a single photon into a certain mode $\mathbf{k}_{out}$ is given by\cite{PhysRevA.66.065403}
\begin{equation}
P\left(\mathbf{k}_{out}\right) \propto \frac{1}{N}\left|\sum_j e^{-i\left(\mathbf{k}_{in}+\mathbf{k}_{c}-\mathbf{k}_{r}-\mathbf{k}_{out}\right) \cdot \mathbf{x}_j}\right|^2.
\end{equation}

Assisted by the collective enhancement of the Rydberg superatom, this emission pattern is highly directional: for mode  $\mathbf{k}_{out}=\mathbf{k}_{in}+\mathbf{k}_{c}-\mathbf{k}_{r}$, each atom shares the same phase, and the emission amplitude is maximally enhanced.
In single photon storage-and-retrieval experiments, the control laser and the read-out laser are often chosen to have the same mode $\mathbf{k}_{c}=\mathbf{k}_{r}$, so the retrieved single photon will propagate along the same mode as the input, $\mathbf{k}_{out}=\mathbf{k}_{in}$.

The efficient matter-light interface makes the Rydberg superatom a promising candidate for photonic quantum memory. Unlike their ground state counterpart, Rydberg quantum memories interact strongly, providing new possibilities for realizing deterministic quantum photonic devices,
such as conditional phase shifter,\cite{Firstenberg2013, doi:10.1126/sciadv.1600036, Thompson2017} entangling gate,\cite{Tiarks2019, PhysRevX.12.021035} and entanglement filter.\cite{ye2023photonic}

Rydberg superatom also differs from a single atom in terms of its coherent properties. As a collective excitation, Rydberg superatom is the superposition of each microstates: $\sum_{j=1}^N e^{i \phi_j}|g_1,g_2,...,r_j,...,g_N\rangle$. Therefore, the coherence of the Rydberg superatom largely depends on how well the relative phases $\phi_j$ between each microstates are maintained.
Any effect that smears the relative phases will cause superatom decoherence.

One of the dominant Rydberg superatom decoherence mechanisms is motional dephasing.~\cite{PhysRevLett.125.263605} After the preparation of the Rydberg superatom, the excitation lasers imprint an effective phase pattern, so-called spin wave, onto the atoms: $\phi_j=\Delta\mathbf{k} \cdot \mathbf{x}_j$. Here $\Delta\mathbf{k}$ is the differential wave vector from the excitation lasers, which corresponds to a spatial phase grating with period $\lambda_{\rm spinwave} = 2 \pi /|\Delta\mathbf{k}|$ along the direction of $\Delta\mathbf{k}$.
For a gas of atoms with mass $m$ and temperature $T$, the atomic speed can be described by the Maxwell-Boltzmann distribution, $f(v)=\left(m / 2 \pi k_B T\right)^{-3 / 2} e^{-m v^2 / 2 k_B T}$.
For an atom with velocity $\mathbf{v}_j$, the motion-induced phase accumulates as $\phi_{j} = \Delta\mathbf{k}\cdot\mathbf{v}_j t$. During the read-out process, the probability $\eta$ of converting the Rydberg superatom into the mode-matched direction is proportional to: $\left|\sum_j f\left(v_j\right) e^{i \phi_j}\right|^2  \sim\left|\int e^{-m v^2 / 2 k_B T} e^{i \Delta k v t} d v\right|^2 \sim e^{-t^2 / \tau^2}$.\cite{berman2011principles} 
Here, the motional dephasing time is $\tau=\sqrt{m / k_B T} / \Delta k=\lambda_{\rm spinwave} /\left(2 \pi \sqrt{k_B T / M}\right)$. 
For two-photon Rydberg excitation schemes, counter-propagating lasers schemes have larger spin wave periods, thus alleviating the motional superatom dephasing.
For $^{87}\rm Rb$,  counter-propagating $780~\rm~nm$-$480~\rm~nm$ excitation scheme has a spin wave period of $\lambda_{\rm spinwave} = 2 \pi /|\Delta\mathbf{k}| = 1.25~\mu \rm m$, leading to a coherence time of $\sim 6~\mu \rm s$, for $10~\mu \rm K$ atomic temperature.

To suppress motional dephasing, an effective technique is to confine the atoms within spatial regions that are much smaller than the spin wave period. For ground states superatom, a one-dimensional optical lattice along the direction of the spin wave can be used to overcome the motional dephasing and extend the coherence time to minutes scale.\cite{PhysRevA.87.031801} However, while red-detuned far-off-resonance optical traps are attractive for ground atoms, {they are generally repulsive for Rydberg atoms.\cite{PhysRevA.72.022347,Adams_2020}} Therefore, traps that could simultaneously confine ground and Rydberg states are essential for extending motional decoherence times. To this end, optical traps with ``magic'' wavelengths are proposed \cite{PhysRevA.84.043408,PhysRevA.72.022347} and demonstrated. \cite{Li2013} In ground-Rydberg state insensitive lattices, the Rydberg superatom storage times and many-body Rabi oscillation coherence times have been extended to tens of microseconds. \cite{PhysRevA.98.033411,PhysRevLett.128.123601}

In addition to the ground-Rydberg-state magic trapping lattice, progress in extending the superatom coherence times was also made using other approaches. Using 297~nm single-photon Rydberg excitation, the quantum state shelving between Rydberg and ground states superatom was demonstrated. \cite{Li2016} The relatively short-lived Rydberg superatom can be coherently transferred to the ground state superatom, whose coherence time can be orders of magnitude longer.
Furthermore, atomic samples prepared at lower temperatures also help to mitigate motional dephasing. \cite{PhysRevA.101.013421}
In addition, it was recently shown that Rydberg spin wave motional dephasing can also be suppressed by introducing another spin wave using off-resonant ac-Stark lattice modulation.\cite{kurzyna2024long}

Due to their collective nature, the coherence properties between two Rydberg superatoms also differ from those of two single atoms. Multiple collective Rydberg excitations (multiple Rydberg superatoms) can be prepared for ensembles separated by a distance larger than the { blockade} radius. 
After the preparation of two Rydberg superatoms, the interaction between the pair of Rydberg atoms leads to the accumulation of phases over time,
\begin{equation}
|RR\rangle \sim \sum_{j, j^{\prime} \neq j}^N e^{i \frac{V_{j j^{\prime}}}{\hbar} t}|g_1, ...,r_j,...,r_{j^{\prime}},...,g_N\rangle.
\end{equation}
Here, $V_{j j^{\prime}}= C_6/x_{j j^{\prime}}^{6}$ is the vdW interaction between the Rydberg atom pair separated by $x_{j j^{\prime}}$. Since the interaction is distance-dependent, the accumulation of random phases over time leads to a collective two-body dephasing.\cite{PhysRevLett.108.030501} To reduce such a dephasing mechanism, the interaction disorder between Rydberg superatoms needs to be suppressed, for instance, by making the superatom size much smaller than the separation between superatoms.

Notably, the Rydberg interaction-induced decoherence mechanism can be harnessed for the effective preparation and manipulation of photonic quantum states.\cite{PhysRevLett.108.030501, PhysRevA.86.021403} 
For example, interaction dephasing-based Rydberg single-photon source, \cite{PhysRevA.106.L051701,PhysRevA.109.013710} non-local quantum optics,\cite{Busche2017} and dissipative entanglement creation\cite{ye2023photonic} have been experimentally demonstrated.

\section{APPLICATIONS IN QUANTUM INFORMATION PROCESSING}
\label{sec:quantuminformation}
\subsection{Quantum logic gates}
\begin{figure}
\centering  
\includegraphics[width=0.9\linewidth]{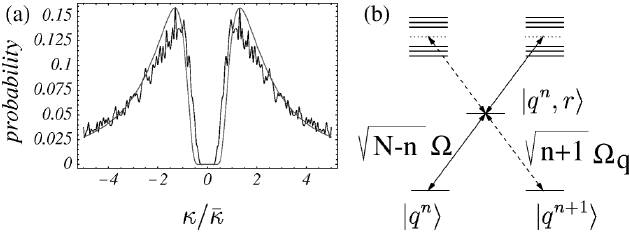}
\caption{\label{fig3}(a) The displayed outcome presents both the results of Monte Carlo simulations and analytical approximations for a random gas in a finite volume, illustrating the probability distribution for frequency shifts of doubly excited Rydberg states. (b) Dipole blockade facilitates the generation of successive Fock states.
Reproduced with permission from 
Lukin {\it et al}., Phys. Rev. Lett. {\bf 87}(3), 037901 (2001). Copyright 2001
American Physical Society.\cite{PhysRevLett.87.037901}}
\end{figure}
The strong and long-range interactions furnish a direct pathway for the realization of two-qubit gates within Rydberg superatoms. {The pioneering gate protocol in Ref.~\onlinecite{PhysRevLett.87.037901}, entails a Rydberg superatom blockade gate.} In an ensemble comprising $N$ identical atoms, each atom has the opportunity to be excited with equal probability, leading to the excitation of symmetric collective states exclusively starting from the ground state. However, the presence of resonant dipole-dipole interactions introduces a spectrum of doubly excited states, each state being separated by an energy gap of {approximately $\hbar\bar{\kappa} = \wp_{rp'}\wp_{rp''}/V$ in ensembles confined within a finite volume $V$, where $r$, $p'$, and $p''$ denote long-lived Rydberg states, $\wp_{rp'}$ and $\wp_{rp''}$ denote the dipole matrix elements for the corresponding transitions.} Should the energy gap exceed the linewidth of the Rydberg state, resonant excitations from singly to doubly excited states experience significant suppression. Figure~\ref{fig3} illustrates the probability distribution for frequency shifts of doubly excited Rydberg states by Monte Carlo simulations of $3\times10^4$ positions within a rectangular box along with the corresponding analytical approximations for a random gas. The strength of the interaction $\kappa$ between any pair of atoms exceeds $\bar{\kappa}$, dominating most of the probability.
This inhibition facilitates the production of collective spin states, enhancing spectroscopic sensitivity beyond the standard quantum limit. It also aids in the preparation of a qubit that exhibits superpositions between the ground state and the single-excited storage state. This capability proves valuable for generating entanglement between two groups of atoms, drawing on principles similar to those described in Ref.~\onlinecite{PhysRevLett.85.2208}.
This regime sets the groundwork for the emergence of superatoms in atomic gases, encompassing the Rydberg dressing, and opens avenues to scalable quantum logic gates within optical or magnetic trap arrays.~\cite{PhysRevLett.105.160404,PhysRevLett.116.243001,PhysRevLett.124.033603} Subsequently, experimental demonstrations of superatom gates have been achieved,~\cite{PhysRevLett.100.253001} along with the proposal of a scheme combining EIT and Rydberg blockade in an atomic ensemble.~\cite{Weatherill_2008}

To streamline operational steps and address practical applications, a single-step multiatom entanglement scheme based on individual atoms within an ensemble has emerged. M\"uller \textit{et al.} proposed a scheme to achieve a high-fidelity gate of many particles using Rydberg interactions and combining EIT,~\cite{PhysRevLett.102.170502} which can be applied to multiparticle interferometry measurements.~\cite{PhysRevLett.93.140408,PhysRevA.75.023615}. {The basic idea is that a control atom conditionally affects the dynamics of spatially separated ensemble atoms through long-range Rydberg interactions. The control atom consists of two ground states $\vert 0_c\rangle$ and $\vert 1_c\rangle$, and a Rydberg state $\vert r_c\rangle$, where states $\vert 1_c\rangle$ and $\vert r_c\rangle$ are resonantly coupled. Ground states $\vert A_e\rangle$ and $\vert B_e\rangle$ of atoms in the ensemble are off-resonantly coupled to intermediate state $\vert P_e\rangle$  (Rabi frequency $\Omega_{p}$), and to Rydberg state $\vert R_e\rangle$ (Rabi frequency $\Omega_c$). Under the EIT condition (two-photon resonance and $\Omega_c\gg \Omega_p$), {the collective $\pi$-pulse evolution can be performed} (e.g. $\vert 0_c\rangle\vert A^N_e\rangle\to \vert 0_c\rangle\vert A^N_e\rangle$) if the control atom is in state $\vert 0_c\rangle$. However, when the control atom is initialized in state $\vert 1_c\rangle$ and excited to $|r_c\rangle$, causing an energy shift of the $\vert R\rangle$ state. Using the resulted Raman transfer between $\vert A\rangle$, and $\vert B\rangle$, the collective operations $\vert 1_c\rangle|A_e^N\rangle$ $\leftrightarrow$ $\vert 1_c\rangle|B_e^N\rangle$, and $\vert 1_c\rangle|B_e^N\rangle$ $\leftrightarrow$ $\vert 1_c\rangle|A_e^N\rangle$ {can be performed}.} Based on this model, the fidelity of the { controlled-${\rm NOT}^N$ logic gate} can reach 99\%.
Similar setting can also be used for nonadiabatic geometric quantum computation to reduce evolution time and mitigate decoherence.~\cite{PhysRevLett.109.170501,PhysRevA.98.032313,Maity2022,PhysRevA.102.042607} In 2018, nonadiabatic geometric quantum computation based on superatoms was proposed,~\cite{PhysRevA.98.032313} and can be extended to the 3-qubit case with two species of superatoms in a two-mode cavity.~\cite{Maity2022} Guo \textit{et al.} utilized dynamical invariant and {zero systematic-error sensitivity methods} to reverse design gate pulses,~\cite{PhysRevA.102.042607} optimizing gate fidelity and robustness against control errors. The average fidelity of the {two-qubit controlled-NOT (CNOT) gate} in their scheme is 0.9971.

For multiatom ensemble qubits, Saffman's group demonstrates coherence control between two ensemble qubits and achieves Rydberg blockade between them with a fidelity of 0.89(1).\cite{PhysRevLett.115.093601} However, fluctuations in the collective Rabi frequency could greatly affect the ensemble qubits operation fidelity. Thus, a scheme is proposed to simulate quantum process tomography to characterize Rydberg gates based on collective states using optimal stimulated Raman adiabatic passage pulses.~\cite{Beterov_2016} Recently, significant progress has been made in achieving entanglement gates through coherent manipulation of deterministically assembled arrays, employing local optical control to achieve entanglement using interatomic interactions, although individual addressing of atoms remains a challenge experimentally.~\cite{PhysRevLett.119.160502,Scholl_2023,Evered2023,ma2023high}
Additionally, due to the excellent transmission characteristics of photons but the lack of strong interactions, entangling gate schemes for photon-atom hybridization have been proposed.~\cite{PhysRevA.85.032309,PhysRevLett.112.040501,Hao2015,PhysRevA.93.040303,PhysRevA.92.030303} Specifically, the combination of a Rydberg blockade and optical cavities effectively enhances the gate robustness,~\cite{PhysRevA.93.040303} while a two-photon interference through virtual photon exchanges renders the internal state dynamics of two neutral-atom qubits insensitive to the thermal photon state.~\cite{PhysRevA.92.030303}

\begin{figure}
\centering  
\includegraphics[width=0.9\linewidth]{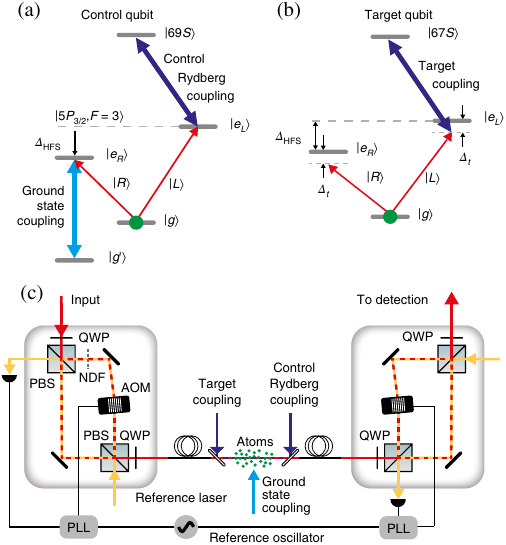}
\caption{\label{figadd1}
Level scheme for EIT storage of the control qubit (a) and target qubit (b). (c) The schematic of the experimental setup. {Here, $|R\rangle$ and $|L\rangle$ are used to define polarization states of photons.}
Reproduced with permission from Tiarks
{\it et al.}, Nature Physics {\bf 15}(2), 124-126 (2019). Copyright 2019 Springer Nature.\cite{Tiarks2019}}
\end{figure}

Tiarks {\it et al.} achieved the first experimental realization of a photon-photon quantum gate using Rydberg interactions, as documented in their study.\cite{Tiarks2019} By harnessing both EIT and Rydberg interactions within an ultracold atomic ensemble, they effectively encoded quantum information carried by photons into specific atomic states, as depicted in Fig.~\ref{figadd1}(a). This mapping facilitates the control of photon loss through post-selection, thereby mitigating its detrimental impact on quantum information and ensuring resilience against errors stemming from photon loss. Notably, the photon interaction occurs exclusively when they share the same polarization $|L\rangle$ due to the Rydberg component, and their interaction during temporal and spatial overlap results in the accumulation of a conditional phase shift, thus realizing a controlled phase flip gate, as illustrated in Fig. \ref{figadd1}(b). By altering the basis, this gate can be converted into a {CNOT} gate. The fidelity of the {CNOT} truth table and the entangling-gate fidelity, post-selected upon the detection of control and target photons, are reported to be 70(8)\% and 63.7(4.5)\%, respectively. This technique employs effectively Rydberg interactions and EIT, allowing for strong, long-range interactions between photons while also solving the difficulty of restricted photon interactions. By mapping Rydberg interactions onto photons, this gate approach stores and manipulates quantum information, demonstrating high fidelity and potential advances in quantum communication systems and quantum networking applications.

\begin{figure}
\centering  
\includegraphics[width=0.9\linewidth]{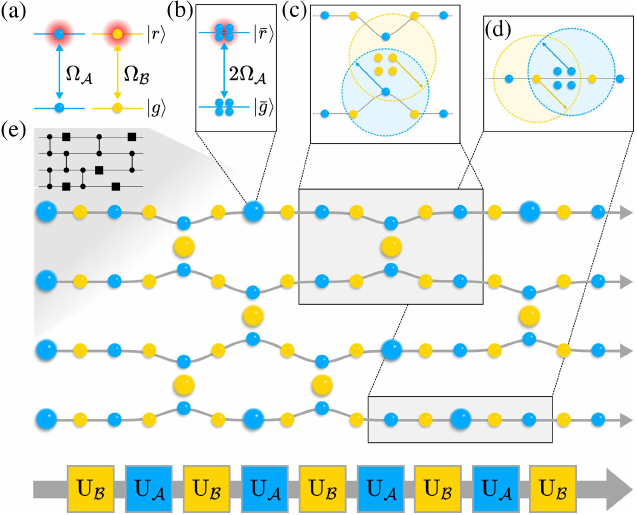}
\caption{\label{fig4}Schematic diagram of converting a circuit into an atomic arrangement, which consists of two types of atoms, and superatoms are used as impurities.
Reproduced with permission from Cesa {\it et al.}, Phys. Rev. Lett. {\bf 131}(17), 170601 (2023). Copyright 2023
American Physical Society.\cite{PhysRevLett.131.170601}}
\end{figure}

Recently, the Rydberg superatoms were utilized in a model for quantum computation using dual-species Rydberg atom arrays,~\cite{PhysRevLett.131.170601} which relies solely on global driving [see Fig.~\ref{fig4}], thus eliminating the requirement for local addressing of the qubits. Ceas and Pichler introduce two distinct constructions: One imprints the circuit within the trap positions of the atoms, executed by pulses; the other encodes the algorithm within the global driving sequence, making the atom arrangement circuit independent. Central to this scheme is the concept of information flow, specifically focusing on the ability of qubits to propagate described by $|\Psi(k+1)\rangle = {\rm U}_{\mathcal{B}}{\rm U}_{\mathcal{A}}{\rm U}_{\mathcal{B}}|\Psi(k)\rangle$, achieved under the conditional occurrence of the Rydberg blockade. Furthermore, the initialization of the circuits must align with the information flow. For the basic requirements of quantum computation, the construction of single- and two-qubit gates requires the assistance of impurity superatoms [see Figs.~\ref{fig4}(c) and \ref{fig4}(d)]. Arbitrary single-qubit gates, including the Hadamard gate, are realized by using superatoms $\mathcal{A}$ within the wires to enhance the Rabi frequency, facilitating collective phase gates $\mathcal{Z}_{\rm tot} = \bigotimes_{Q}{\rm Z}_{Q}$. Additionally, for two-qubit entangling gates, another superatom $\mathcal{B}$ placed between the wires can induce a $\rm CZ$ gate. The study demonstrates that a quadratic increase in atom number is sufficient to eliminate the need for local control, thereby achieving a universal quantum processor. The authors also provide explicit protocols for all stages of arbitrary quantum computations and discuss error suppression strategies tailored to their model. Although the scheme is grounded in dual-species processors adhering to Rydberg blockade constraints, it holds potential for adaptation to alternative setups. This study opens new avenues for universal quantum computation with Rydberg superatoms.

\subsection{Quantum entanglement}
Quantum entanglement plays a crucial role in quantum computing and quantum information processing. Early experiments of entangled states have focused on photons and ions.~\cite{PhysRevLett.103.020503,PhysRevLett.103.020504,Häffner2005} However, these schemes face challenges in scaling to multiqubit entanglement and have a low success probability. Rydberg ensembles offer a promising solution to overcome these limitations and realize scalable entanglement of neutral atoms. Atom vapors serve not only as quantum memory, but also as ``good'' qubits due to strong Rydberg interactions. { The theoretical study shows that high-efficiency cluster states can be created via single-qubit measurements with Rydberg ensembles.}~\cite{PhysRevA.79.022304} Utilizing a fast entanglement generation scheme with a Lipkin-Meshkov-Glick-type Hamiltonian, a 700-atom cat state has been generated with a 50-fold improvement in the size of the entangled state.~\cite{PhysRevA.98.043836} To further enhance fidelity, G\"{a}rttner proposes a scheme to herald the entangled state by detecting an ion, achieving near-unity fidelity.~\cite{PhysRevA.92.013629} Additionally, addressing decoherence concerns, one can combine Zeno dynamics and shortcuts to adiabaticity with Rydberg superatoms, encoding quantum information in the collective state of superatoms to rapidly realize entangled states, including Bell states and $W$ states.~\cite{Ji2020} Similarly, fast conversion of three-particle Dicke states to four-particle Dicke states with Rydberg superatoms can be implemented based on Zeno dynamics and shortcuts to adiabaticity, offering high fidelity and robustness against decoherence.~\cite{https://doi.org/10.1002/qute.202200173} In Rydberg ensembles, large-scale entangled states can also be prepared {theoretically} by sending two Rydberg superatoms belonging to two different atomic $W$ states into a cavity without qubit loss.~\cite{Zhang_2022}

\begin{figure*}
\centering  
\includegraphics[width=0.9\linewidth]{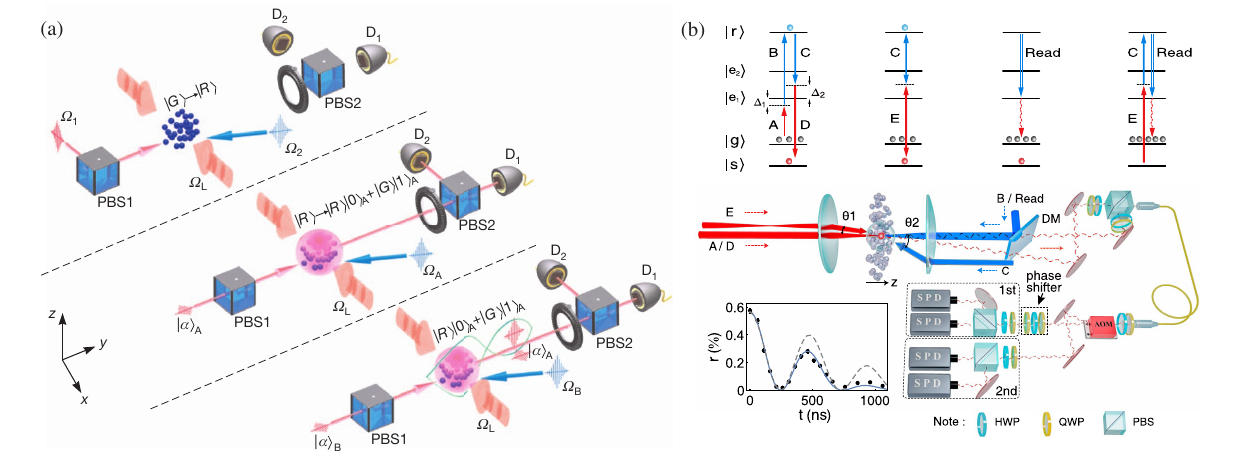}
\caption{\label{su_3_4} Entanglement between Rydberg superatom and photon. (a) Schematic diagram of the entanglement of light with optical atomic excitation. Mapping the collective atomic excitation into quantum fields $|\phi\rangle_{A,B}$ by implementing resonance on the $|r\rangle$ and $|e\rangle$ used laser fields $\Omega_{A,B}$.
Reproduced with permission from Li
{\it et al.}, Nature {\bf 498}(7455), 466-469 (2013). Copyright 2013 Springer Nature.\cite{Li2013} (b) Experimental preparation of semideterministic atom-photon entanglement.
Reproduced with permission from Li {\it et al.}, Phys. Rev. Lett. {\bf 123}(14), 140504 (2019). Copyright 2019
American Physical Society.\cite{PhysRevLett.123.140504}}
\end{figure*}

Photons are excellent carriers for quantum information, and the entanglement between atom and photon holds promise for achieving long-range quantum information distribution. Most ensemble-based atom-photon entanglement schemes are probabilistic and suffer from limited scalability. In 2013, Li {\it et al.} introduced a new protocol to generate entanglement between light and optical atomic excitations with intrinsic determinism.~\cite{Li2013} In this scheme, the dephasing of optical atomic coherence is suppressed by state-insensitive confinement of ground and Rydberg states of ultra-cold atomic gas trapped in a one-dimensional optical lattice at 1004~nm. This confinement eliminates the differential energy shift between the ground state and the Rydberg state $\delta U=U_{r}(r)-U_{g}(r)$ and maintains coherence between them. Such a scheme has significant implications for extending the ground-Rydberg coherence and building functional Rydberg quantum networks. Figure~\ref{su_3_4}(a) illustrates the three-step process for the entanglement protocol: (i) Excitation of the atomic ensembles from the ground state $|G\rangle$ to the single excited Rydberg state $|R\rangle$. (ii) Generation of the entanglement state $|R\rangle\rightarrow|R\rangle|0\rangle_{A}+|G\rangle|1\rangle_{A}$ through a retrieval field $\Omega_{A}$. (iii) Verification of the atom-photon entanglement by analyzing the correlations of photoelectric detection events. This scheme is compatible with quantum logic operations and long-term storage of quantum information, paving the way for quantum networks that enable secure communication and distributed quantum computing.

In 2019, an alternative atom-light entanglement protocol was experimentally demonstrated, leveraging Rydberg blockade and Raman coupling techniques. \cite{PhysRevLett.123.140504} The manipulation of atoms within a confined volume of approximately 10 $\mu$m in all three dimensions ensures high effectiveness of Rydberg blockade, with an atomic density of approximately $10^{11}/\rm cm^3$. In the schematic depicted in Fig.~\ref{su_3_4}(b), the atomic ensemble is initially prepared in a ground state $|g\rangle$, following which Rydberg blockade is employed to sequentially generate two collective excitations—one in a ground state $|s\rangle$ and the other in a Rydberg state $|r\rangle$. These collective excitation states, referred to as two superatoms, possess distinct momenta,  are then entangled by incorporating the momentum degree of freedom and interfering them through Raman coupling. Finally, one excitation is retrieved using a read beam, converting the momentum-entangled atomic state into entanglement between the polarization of the single photon and the momentum of the remaining atomic excitation, thereby achieving atom-photon entanglement. 
The high fidelity measurement of entanglement between the polarization of the single photon and the momentum of the atomic excitation underscores the practicality of employing this approach for quantum information processing tasks like quantum repeaters and networks.
Although this scheme is semi-deterministic compared to the approach in Ref. \onlinecite{Li2013}, it remains immune to phase fluctuations and is devoid of high-order or vacuum components, which are essential for remote entanglement generation. 
After that, the same group presented a report detailing the deterministic creation of entanglement between an atomic ensemble and a single photon.\cite{PhysRevLett.128.060502} This achievement was accomplished through a cyclical process termed ``retrieving and patching'', which was applied sequentially to the Rydberg levels. By leveraging the Rydberg blockade effect, they ensured that a full excitation was generated only if a preceding excitation existed in one of the Rydberg levels. This process led to the creation of a photon in a delayed temporal mode, entangled with the Rydberg levels hosting the collective excitation. Verification of the entanglement between the atomic ensemble and the single photon was conducted by retrieving the atomic excitation as a second photon and performing correlation measurements. The results yielded an entanglement fidelity of 87.8\%.

The entanglement between photons holds tremendous potential across various fields, such as metrology and quantum computing. { A challenge is that interactions between photons are inherently weak. A lot of efforts have been spent on creating sizable photon-photon interactions}.~\cite{Heisenberg1936,Busche2017,Firstenberg2013} Numerous schemes have been proposed for the preparation of multiphoton entanglement, including spontaneous parametric downconversion~\cite{RevModPhys.84.777,PhysRevLett.121.250505} and the entangling of single photons through atom-photon interactions.~\cite{PhysRevA.58.R2627,PhysRevLett.95.110503} However, these schemes often suffer from drawbacks such as poor scalability of the photon number, complex setups, and other requirements on the system. Recently, Yang \textit{et al.} proposed a novel scheme to generate multiphoton entanglement with a Rydberg superatom.~\cite{Yang2022} In their approach, an atomic ensemble is used, where an atomic entanglement state $|\Psi\rangle_\mathrm{a}$ is prepared by applying collective pulses with the Rydberg blockade mechanism. Subsequently, a ``retrieving and patching'' sequence is applied to $|r_1\rangle$, causing the Rydberg superatom to emit a photon and recreate the excitation. This sequence is repeated for $|r_2\rangle$, {resulting in the creation of the entanglement state $(|0\rangle_1|1\rangle_2|L\rangle+|1\rangle_1|0\rangle_2|E\rangle)/\sqrt{2}$. By repeating these processes $(m-1)$ times and retrieving the two atomic qubits, the GHZ-type multiphoton entanglement $(|L\rangle^{\otimes m}+|E\rangle^{\otimes m})/\sqrt{2}$ can be obtained, where $|L\rangle$ and $|E\rangle$ represent a photon in the late and early modes, respectively}. To enhance the performance of multiphoton entanglement, the scheme employs a low-finesse cavity to improve photon collective efficiency and polarizing beamsplitters with a Pockels cell in measurement to minimize exposure. The fidelity of the final prepared three- and six-photon GHZ states is $82.9\pm0.3\%$ and $61.8\pm2.6\%$, respectively.

\begin{figure}
\centering  
\includegraphics[width=0.9\linewidth]{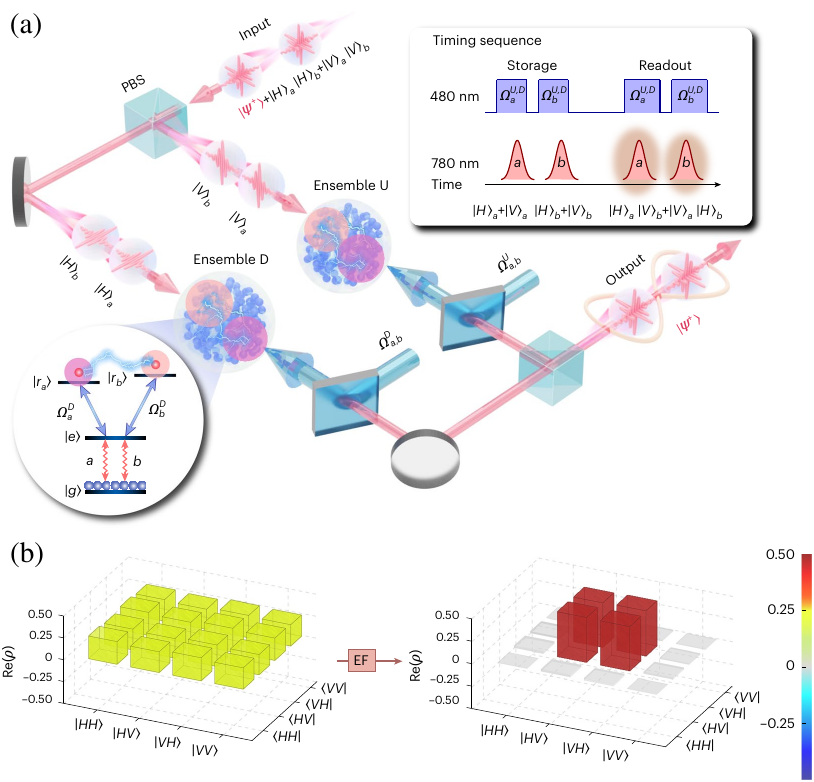}
\caption{\label{EF} Photonic entanglement filter with Rydberg superatoms. (a) Illustration of Rydberg entanglement filter protocol. (b) Extracting high-fidelity entangled state from input state with low fidelity.
Reproduced with permission from Ye {\it et al.}, Nature Photonics {\bf 17}(6), 538–543 (2023). Copyright 2023 Springer Nature.\cite{ye2023photonic}}
\end{figure}

Strongly interacting Rydberg superatoms play a crucial role not only in generating entanglement but also in enabling deterministic manipulation of entangled photonic states.
In Ref.~\onlinecite{ye2023photonic}, a novel photonic entanglement filter was realized using two Rydberg superatoms.
The demonstrated entanglement filter transmits the desired photonic entangled state and blocks unwanted ones.
As illustrated in Fig.~\ref{EF}(a), to achieve quantum state filtering, the undesired photonic components, $|HH\rangle$ and $|VV\rangle$ states, are converted into double Rydberg excitations in the same atomic ensemble via EIT photon storage.
These double excitations can be eliminated using either Rydberg blockade or interaction-induced dephasing. On the other hand, the target entangled state, $|H\rangle|V\rangle+|V\rangle|H\rangle$ is protected in a decoherence-free subspace and can pass through the entanglement filter.
Compared to prior repeat-until-success approaches based on probabilistic linear-optical systems, the Rydberg entanglement filter harnesses the strong atomic interaction and is, therefore, inherently deterministic. As shown in Fig.~\ref{EF}(b),
photonic entanglement with fidelity $\sim 99 \% $ can be extracted from an input state even when the initial fidelity is low.
Such an entanglement filter opens new possibilities for scalable photonic quantum information processing with Rydberg superatom arrays.

The overall outlook for creating quantum entanglement using Rydberg superatoms is very positive. Areas including quantum computing, quantum communication, and basic quantum physics stand to benefit from further research and development in this field. These upgrades might pave the way for ground-breaking discoveries with far-reaching consequences in various science and technology areas. 

\subsection{Quantum simulation}
Quantum simulation serves as a powerful tool for understanding and predicting many-body dynamics, allowing access to a wide range of phenomena, including thermal equilibrium relaxation in closed systems {and non-thermal behaviors in integrable and disordered systems.\cite{Daley2022}} 
A particularly promising platform for investigating the dynamics of strongly interacting many-body quantum systems is provided by coherent laser-driven Rydberg atomic gases.\cite{PhysRevA.91.063401,PhysRevLett.115.125301,dingErgodicityBreakingRydberg2024} Currently, spin systems based on Rydberg atoms have emerged as a promising platform for quantum simulation because of their superior interaction flexibility and strength, high controllability, and effective isolation from the environment.

One {theoretical} strategy suggested by Sun and Robicheaux involves simulating Rydberg gases and exploring two-body correlation by driving the superatoms, which can be subdivided into smaller regions called pseudoatoms, with a laser having a Rabi frequency following a Poissonian distribution. \cite{Sun_2008} This method incorporates the classical correlation\cite{PhysRevA.74.052110} and provides a more accurate depiction of the Rydberg gas compared to the mean field approximation.
Using the superatom picture, Olmos and colleagues conducted extensive {theoretical} research on fermionic collective excitations and thermalization in a symmetric quasi-one-dimensional ring lattice gas of Rydberg atoms.~\cite{PhysRevLett.103.185302,Olmos_2010}
In the {theoretical} studies of Refs.~\onlinecite{PhysRevLett.104.043002,Bijnen_2011}, the Rydberg superatoms serve as building blocks for the creation of Rydberg crystals using tailored laser excitation schemes. These crystals are used as representations of dilute metallic solids with tunable lattice parameters, allowing access to a wide range of fundamental phenomena in condensed matter physics.

\begin{figure}
\centering
\includegraphics[width=0.9\linewidth]{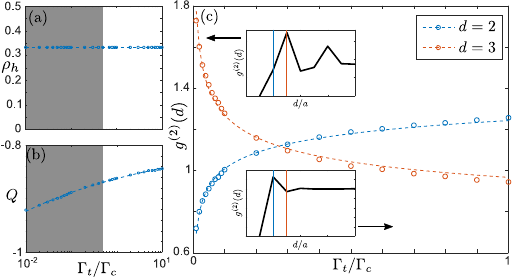}
\caption {\label{tu8.png}Average density of holes $\rho _{h}$,  Mandel $Q$ parameter for the total number of holes, and amplitudes of the density–density correlation function $g^{\left ( 2 \right ) } \left ( d \right ) $ for the periods of $d = 2a$ (blue circles) and $d = 3a$ (red circles), versus the hopping rate $\Gamma _{t}$.
 Reproduced with permission from 
Letscher {\it et al}.,  New J. Phys. {\bf 19}(11), 113014 (2017). Copyright 2017
Institute of Physics, licensed under a Creative Commons Attribution 3.0 licence.\cite{Letscher_2017} }
\end{figure}

In 2017, researchers explored {a numerical
study on} the intricate dynamics of holes within a chain of Rydberg superatoms, exploring their behavior under the influence of driving forces and dissipation.~\cite{Letscher_2017} The study introduces an effective model depicting these holes as hard rods with a length of 2$a$, providing insights into their creation, annihilation, and spatial correlations within the lattice. These holes stem from the spontaneous decay of excited superatoms and can be annihilated when two holes are separated by an excited superatom, thereby influencing the equilibrium dynamics of the system. The steady-state distribution of holes converges to a density of 1/3 when the detuning of the laser matches the interaction strength between neighboring superatoms, showcasing the intricate interplay between excitation processes and decay mechanisms.
Spatial correlations among holes exhibit liquid-like behavior, characterized by a length scale of 2$a$ and decay over distances comparable to the lattice constant. An effective model for describing holes is established, offering an intuitive explanation by examining the influence of the hopping rate on the many-body steady state. As depicted in Fig.~\ref{tu8.png}, increasing the hopping rate $\Gamma_t$ renders the holes mobile, gradually reducing the peak of the correlation function $g^2(d)$ at $d=3a$, while for a large hopping rate, the correlation function $g^2(d)$ exhibits a period $d=2a$. The Mandel $Q$ parameter describes the number fluctuations of holes, indicating significantly suppressed fluctuations attributed to pair annihilation processes.
These findings yield valuable insights into the collective behavior of holes within the driven, dissipative spin chain of Rydberg superatoms, highlighting the pivotal role of hole dynamics in shaping the overall equilibrium phase of the system. The study contributes to the understanding of many-body dynamics in quantum systems, offering a comprehensive analysis of hole interactions and their implications for the spin dynamics of Rydberg superatom chains.

\begin{figure}
\centering  
\includegraphics[width=0.9\linewidth]{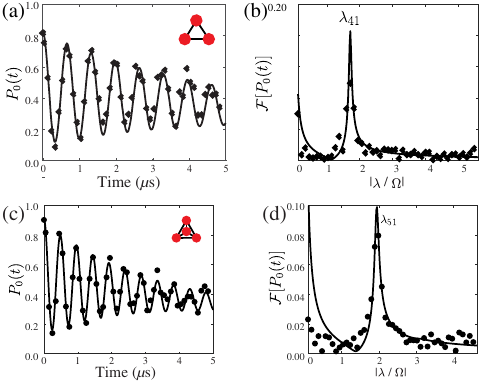}
\caption{\label{tu7}Ground state population and corresponding Fourier transform in $N$-body ($N=3,4$) configuration.
Reproduced with permission from 
 Kim {\it et al}., PRX Quantum {\bf 1}(2), 020323 (2020). Copyright 2020
American Physical Society, licensed under a Creative Commons Attribution 4.0 International license.\cite{PRXQuantum.1.020323}}
\end{figure}

The scheme outlined in Ref.~\onlinecite{PRXQuantum.1.020323} suggests the continuous tuning of the quantum Ising Hamiltonian of Rydberg atoms arranged in three dimensions. In this arrangement, the atoms and pairs of Rydberg blockaded atoms are depicted as vertices and edges, respectively, as illustrated by the geometric red configuration in Fig.~\ref{tu7}. Consequently, various connected graphs of $N$ body configurations of Rydberg atoms are constructed, and their eigenenergies, along with different geometric intermediates, are explored during the structural transformation process. The ground state population and the corresponding Fourier transform of the configurations in the superatom regime are depicted in Fig.~\ref{tu7} for $N=3,4$. Remarkably, the experimentally measured graph-dependent eigenspectra (represented by filled circles) align well with the model calculations of the few-body quantum Ising Hamiltonian (illustrated by lines). This high-dimensional programming of qubit connectivity demonstrated in the study represents a significant step toward the application of programmable quantum simulators.\cite{PRXQuantum.3.030305,PhysRevResearch.5.043037}

\begin{figure}
\includegraphics[width=0.9\linewidth]{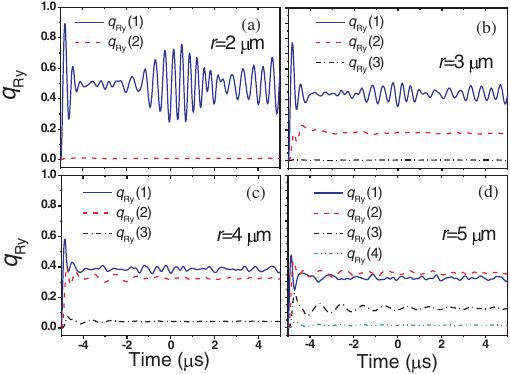}
\caption {\label{fig7.png}The numerically calculated probabilities $q_{Ry}\left(n\right)$ to
excite $n$ Rydberg atoms in the mesoscopic atomic ensemble with $\bar{N}=7$ atoms, randomly distributed in the optical dipole trap with different radius $r$. Reproduced with permission from Beterov {\it et al}., Phys. Rev. A {\bf 90}(4), 043413 (2014). Copyright 2014 American Physical Society.\cite{PhysRevA.90.043413}}
\end{figure}

One of the challenges in observing the Rydberg blockade is to obtain a precise measurement of the quantity of Rydberg atoms. Due to low detection efficiencies, accurately determining the number of excited Rydberg atoms becomes problematic, which is essential for exploiting the Rydberg blockade in quantum information applications. Beterov \textit{et al.} investigated the complex dynamics of Rydberg excitations in atomic ensembles influenced by a classical electromagnetic field.\cite{PhysRevA.90.043413} They demonstrated the observation of collapses and revivals in the average count of Rydberg excitations, resembling the behavior predicted by the Jaynes-Cummings (JC) model. Through the analysis of randomly loaded optical dipole traps, they predicted the occurrence of collapses and revivals in Rabi oscillations, emphasizing how the oscillation frequency depends on the number of interacting atoms. Moreover, the study examines the impact of finite interaction strengths and laser linewidth on the detectability of the revivals. Considering a randomly loaded optical dipole trap with an average atom number of $\bar{N}=7$, data in Fig.~\ref{fig7.png} illustrate the excitation dynamics in a Rydberg state in the ensemble. In Fig.~\ref{fig7.png}(a), the simulated dynamics of the probability of single-atom Rydberg excitation $q_{Ry}\left(1\right)$ for $r=2~\mu\rm m$ show collapses and revivals, while $q_{Ry}\left(2\right)$ approaches zero, indicating an effective Rydberg blockade. However, with increasing radius of the optical dipole trap, the revivals in $q_{Ry}\left(1\right)$ are diminished, and the probability of exciting two Rydberg atoms $q_{Ry}\left(2\right)$ increases, suggesting the breakdown of the Rydberg blockade. These results provide valuable insights into Rydberg blockade phenomena, reducing the need for precise quantification of Rydberg atoms and offering potential applications in quantum information processing and the investigation of sub-Poissonian atom-number fluctuations in mesoscopic atomic ensembles.

{It is noted that significant efforts have been directed towards quantum simulation utilizing Rydberg atom arrays, driven by their exceptional controllability and versatility.\cite{browaeys2020many, kaufman2021quantum} These systems facilitate precise manipulation of individual atoms through laser fields, allowing for the creation of customized potential landscapes and interaction patterns.\cite{doi:10.1126/science.aax9743,labuhn2016tunable} Furthermore, the strong and adjustable interactions among Rydberg atoms enable the emulation of various quantum phase transitions.\cite{de2019observation,kanungo2022realizing,chen2023continuous,bernien2017probing,keesling2019quantum} The scalability of these arrays, capable of accommodating hundreds of and even moreatoms,\cite{ebadi2021quantum,scholl2021quantum,doi:10.1126/science.abg2530} offers researchers a robust platform for exploring large quantum systems and addressing computationally challenging problems.\cite{doi:10.1126/science.aah3778, doi:10.1126/science.aah3752,   Glaetzle2017, Barredo2018Synthetic, doi:10.1126/science.aaw4150,   Bekenstein2020,    doi:10.1126/science.abi8794, doi:10.1126/science.abo6587, wu2022erasure,  Bluvstein2022A, Kim2022, Scholl_2023,   Schine2022Long, Yan2022Quantum,  ORourke2023Entanglement, Srakaew2023A, Bornet2023Scalable,  Chen2024Strongly} Leveraging unique properties of Rydberg superatoms, we anticipate opportunities for simulating quantum spin models, probing topological phases, and studying thermalization. This can be achieved using techniques such as trapping many atoms per blockade radius in optical lattices or arrays of microscopic dipole traps spaced a few micrometers apart.\cite{Côté_2006,weimer2010rydberg,PhysRevLett.107.060402,PhysRevLett.107.263001,weber2015mesoscopic,doi:10.1126/science.aal3837,Whitlock_2017,PhysRevLett.128.123601}}

\section{APPLICATIOINS IN QUANTUM OPTICS}
\label{sec:quantumoptics}
In this section, we first present a brief review of recent experimental and theoretical progress focusing on the nonlocal nonlinear responses, as induced by the dipole-dipole interactions and related to the dipole
blockade effect, of cold Rydberg atoms driven into the EIT regime.\cite{RevModPhys.77.633,Finkelstein_2023} Then
we look back at some important proposals and schemes for achieving efficient nonlinear quantum optics\cite{chang2014quantum}
by considering hybrid systems with Rydberg atoms restricted in cavities, placed
close to waveguides, or trapped above atomic chips.

\begin{figure}
\includegraphics[width=0.9\linewidth]{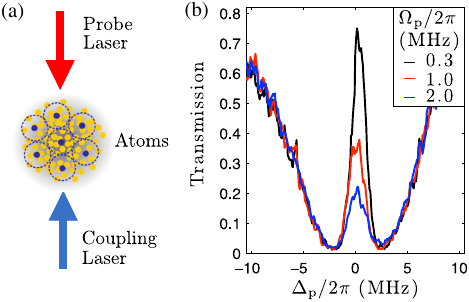}
\caption{\label{wufig1} (a) Schematic of a Rydberg EIT experiment with cold $^{87}$Rb atoms
driven by a probe field on the lower $|g\rangle\leftrightarrow|e\rangle$ transition and a 
coupling field on the upper $|e\rangle\leftrightarrow|r\rangle$ transition in the ladder
configuration. (b) Nonlinear EIT spectra of the probe field where the transmission peak on
resonance ($\Delta_{p}=0$) reduces as probe Rabi frequency $\Omega_{p}$ increases. The
experiment is done for the $|r\rangle=|60S_{1/2}\rangle$ state at atomic
density $\rho=1.2\pm0.1\times10^{10}$ cm$^{-3}$.
Reproduced with permission from Pritchard {\it et al}., Phys. Rev. Lett. {\bf 105}(19), 193603 (2010). Copyright 2010
American Physical Society.\cite{PhysRevLett.105.193603}}
\end{figure}

\begin{figure*}
\includegraphics[width=0.9\linewidth]{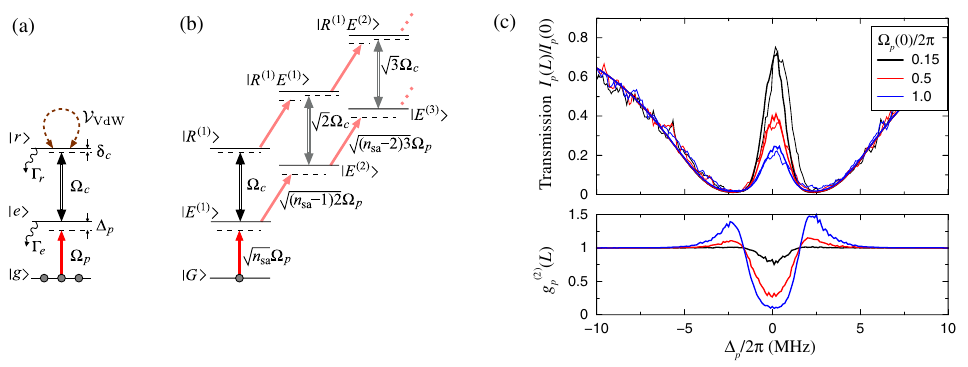}
\caption{\label{wufig2} (a) A ladder scheme of three-level atoms that interact with a probe and a coupling field.
$\mathcal{V}_{\rm vdW}$ denotes the vdW interaction between two atoms in the Rydberg state $|r\rangle$.
(b) Truncated level scheme of a superatom composed of $n_{SA}$ atoms, with collectively enhanced transition
amplitudes involving different numbers of probe and coupling photons. (c) Top: Probe field transmission against detuning $\Delta_{p}$ for different input probe Rabi frequencies. The thin lines are experimental curves
from Ref.~\onlinecite{PhysRevLett.105.193603} while thick lines are numerical simulations based on the superatom model. Bottom: Corresponding intensity correlation function of the transmitted probe field {at the exit from the medium}.
Reproduced with permission from Petrosyan {\it et al.}, Phys. Rev. Lett. {\bf 107}(21), 213601 (2011). Copyright 2011
American Physical Society.\cite{PhysRevLett.107.213601}}
\end{figure*}

\begin{figure}
\includegraphics[width=0.9\linewidth]{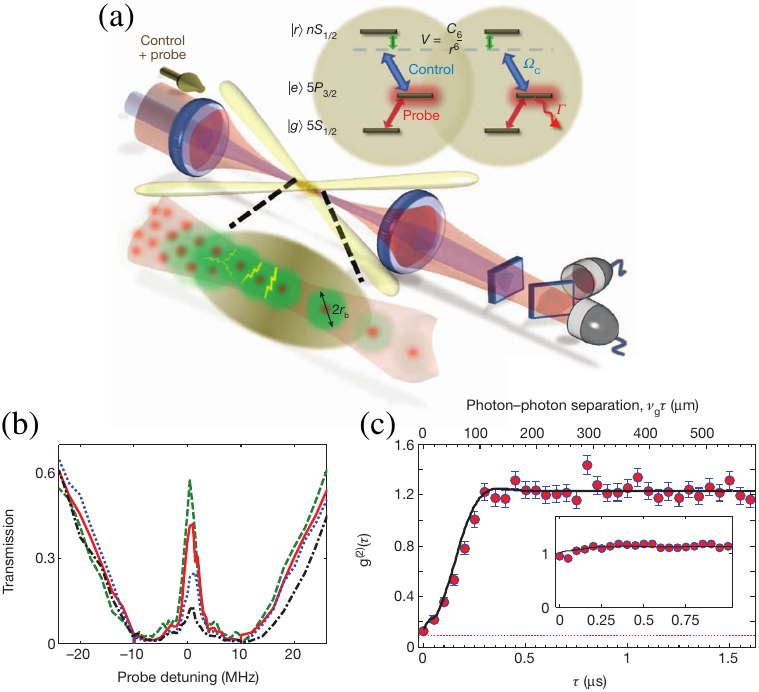}
\caption{\label{wufignew} (a) Rydberg-blockade-mediated interaction between slow photons. (b)
 Transmission versus probe
detuning at various incoming photon rates.
(c) Data points show photon–photon correlation function $g^{(2)}
(\tau)$ at
EIT resonance.
Reproduced with permission
from Peyronel {\it et al}., Nature {\bf488}(7409), 57-60 (2012). Copyright 2012 Springer Nature.\cite{peyronel2012quantum}}
\end{figure}

\subsection{Nonlocal nonlinear optics}
Typical EIT experiments are performed with atoms in low-lying states driven into the three-level $\Lambda$ configuration
{involving an intermediate short-lived excited state and two long-lived ground states.\cite{PhysRevLett.64.1107,PhysRevLett.66.2593}}
When a ground state is replaced with a Rydberg state, it is viable to realize an intriguing transfer
from the linear EIT in the $\Lambda$ configuration to the nonlinear EIT in the ladder configuration,
as a result of both long-lifetime and strong-interaction features of highly excited Rydberg atoms.

In 2010, Adams' group observed in $^{87}$Rb atoms a cooperative optical nonlinearity
arising from Rydberg interactions and manifesting itself as a gradual reduction of the probe transmission
in the EIT window when the probe intensity increases\cite{PhysRevLett.105.193603} as shown in Fig.~\ref{wufig1}.
{The study was performed with an EIT ladder} system involving ground $|g\rangle$, intermediate $|e\rangle$,
and upper $|r\rangle$ levels and {was} explained by a simple two-atom model where one atom contributes to the
three-level EIT system while the other acts as a two-level absorbing system due to dipole blockade.
This fascinating experiment soon {was} better explained by a mean-field superatom model where the collective states
of $n_{SA}=\frac{4}{3}\pi\rho R_{SA}^{3}$ atoms within a blockade sphere of radius
$R_{SA}=(C_{6}\gamma_{e}/\Omega_{c}^{2})^{1/6}$ {defined as the interatomic
distance at which the vdWs interaction equals the
half-width of steady-state population of a single atomic Rydberg state,}\cite{PhysRevA.88.033422} are considered but appropriately truncated\cite{PhysRevLett.107.213601}
as shown in Figs.~\ref{wufig2}(a) and \ref{wufig2}(b). This truncation is done based on two facts: i.e.,
(\emph{i}) at most one atom is allowed to be in the upper Rydberg state and (\emph{ii}) the excitation
to the intermediate state $|e\rangle$ is well suppressed due to quantum destructive interference. Then it is appropriate
to consider only the three lowest-order collective states $|G\rangle=|g_{1},g_{2},...,g_{n_{SA}}\rangle$,
$|E^{(1)}\rangle=\frac{1}{\sqrt{n_{SA}}}\sum_{j}^{n_{SA}}|g_{1},...,e_{j},...g_{n_{SA}}\rangle$, and
$|R^{(1)}\rangle=\frac{1}{\sqrt{n_{SA}}}\sum_{j}^{n_{SA}}|g_{1},...,r_{j},...g_{n_{SA}}\rangle$ for each superatom as
one attempts to estimate the Rydberg excitation probability $\Sigma_{RR}$. With $\Sigma_{RR}$, it is easy to calculate
the conditional polarizability $\alpha=\Sigma_{RR}\alpha_{TLA}+(1-\Sigma_{RR})\alpha_{EIT}$ with $\alpha_{TLA}$
and $\alpha_{EIT}$ being the polarizabilities for the two-level absorbing and ladder EIT systems, jointly
determining the probe field propagation in the Rydberg medium. Such a computationally efficient model is in
quantitative agreement with the experimental results and also predicts in Fig.~\ref{wufig2}(c) a remarkable
antibunching effect in terms of the intensity correlation function $g^{(2)}_{p}(L)$ {obtained}
at {the exit from the medium}, accompanied by a transmission reduction due to cooperative optical nonlinearity.

Through the utilization of Rydberg EIT, Vuleti{\'{c}} and his colleagues showcased their ability to manipulate light fields at the level of single quanta. In their work Ref.~\onlinecite{peyronel2012quantum}, strongly interacting Rydberg states play a crucial role in generating quantum nonlinearity via the photon-photon blockade mechanism, as shown in Fig.~\ref{wufignew}. This blockade prevents the transmission of multi-photon states by precluding the simultaneous excitation of two Rydberg atoms within the blockade radius $r_b$. When an incident single photon is converted into a Rydberg polariton inside the medium under EIT conditions, the Rydberg blockade prevents a second polariton from traveling within the blockade radius of the first one, leading to the destruction of EIT. The interaction between Rydberg atoms creates a strong nonlinearity that results in the attenuation of a second photon approaching the single Rydberg polariton. This phenomenon is based on the physical principle that when the blockade radius exceeds the resonant attenuation length of the medium in the absence of EIT, two photons in a tightly focused beam cannot pass through each other or propagate close to each other inside the medium. By utilizing Rydberg states of $|100S_{1/2}\rangle$, the researchers were able to realize blockade radii $r_b=13~\mu\rm m$, effectively creating a quantum nonlinear absorption filter that converts incident coherent field into a train of non-classical, anti-bunched light pulses. The demonstration of quantum-by-quantum control over light fields in this research opens up new possibilities for non-linear quantum optics based on interacting Rydberg atoms.

Soon afterward, higher-order collective states {were} included in an improved superatom model by going beyond the weak-probe
approximation, which just results in quantitative differences, especially as the probe and coupling fields
are off two-photon resonance, yet without changing main conclusions.\cite{PhysRevA.89.033839} Such a superatom model has
been further extended to realize the polarization-selective nonlinearity and cooperative nonlinear grating unattainable in normal atoms.\cite{PhysRevA.92.063805,Liu:16,PhysRevA.94.033823}
A semianalytical rate-equation model has also been developed for revealing a universal relation between the nonlinear
optical susceptibility and the Rydberg excitation density as interatomic coherence is well suppressed due to strong
dephasing effects or very different Rabi frequencies of the probe and coupling fields.\cite{PhysRevA.89.063407} Moreover,
we note that a many-body iterative approach has been developed to account for the interparticle Rydberg correlations,
though it is applicable only for dilute samples with less than one atom inside a superatom so that Rydberg interactions between
different superatoms become dominant to reduce the many-body computational complexity. This alternative approach allows
one to separate a local linear susceptibility and a nonlocal nonlinear susceptibility comparable in magnitude, hence
achieving huge nonlocal optical nonlinearities beneficial, e.g., to largely enhance the Goos-Hanchen lateral shift and
correlated biphoton generation.\cite{Bai:16,Bai:19,PhysRevA.106.043119,Zhao:23}

\begin{figure}
\includegraphics[width=0.9\linewidth]{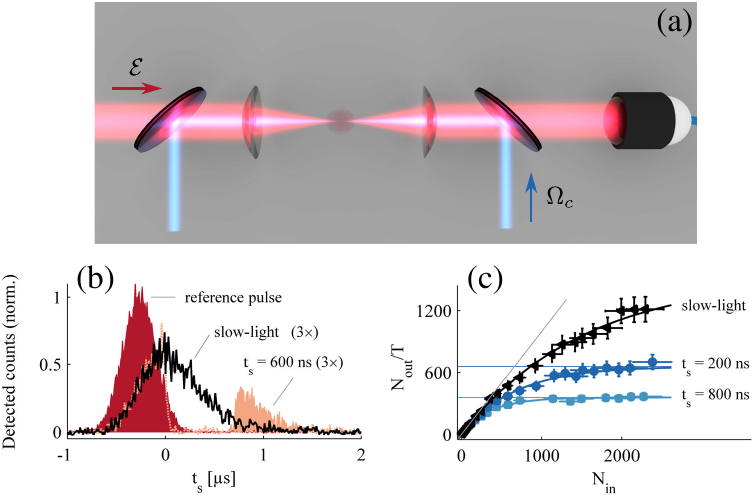}
\caption{\label{wufig4} (a) A light storage and retrieval experiment based on the nonlinear Rydberg EIT technique.
(b) Normalized and background subtracted counts of an input probe pulse when slowly propagating or
stored for $t_{s}=600$ ns. (c) Output photon number normalized by linear process efficiency $T$ against input
photon number for the slow-light case and two storage times. Straight lines represent the linear behavior
(oblique) and the saturation level (horizontal).
Reproduced with permission from  Distante {\it et al}., Phys. Rev. Lett. {\bf 117}(11), 113001 (2016). Copyright 2016
American Physical Society.\cite{PhysRevLett.117.113001}}
\end{figure}
An important application of the linear EIT in normal atoms is to realize the dynamically reversible storage and retrieval of a (classical or quantum) probe pulse by adiabatically switching off and on a control field. Such an important coherent control of the slow-light propagation and storage dynamics has been extended to Rydberg atoms exhibiting nonlocal nonlinear EIT responses. Many experiments show that storing an input probe pulse as a collective Rydberg excitation strongly enhances the Rydberg-mediated interaction compared to that in the slow-light dark-state polariton case. For example,  Distante {\it et al.} observed a strong enhancement of interatomic interactions due to the largely reduced polariton group velocity for short storage times while a weak enhancement dominated by Rydberg-induced dephasing of the polariton's multiparticle components for longer storage times.\cite{PhysRevLett.117.113001} This conclusion is supported by Fig.~\ref{wufig4} where the output photon number $N_{\rm out}$ depends on the input photon number $N_{\rm in}$ clearly in a nonlinear way and becomes saturated for a
sufficiently large $N_{\rm in}$ as described by the input-output relation $N_{\rm out}=N_{\rm max}T(1-e^{-N_{\rm in}/N_{\rm max}})$ with $T$ being the linear process efficiency for small $N_{\rm in}$. This nonlinear light storage experiment can be well explained by extending the superatom model from its steady-state solutions to the dynamic evolution case~\cite{PhysRevA.97.043811} which further predicts an inhomogeneous nonclassical antibunching feature of the retrieved light signal. Another effective method has also been developed to reveal the correlated photon dynamics in dissipative Rydberg media valid in the limit of a large optical depth per blockade radius.\cite{PhysRevLett.110.153601, PhysRevLett.119.043602}

\begin{figure*}
\includegraphics[width=0.95\linewidth]{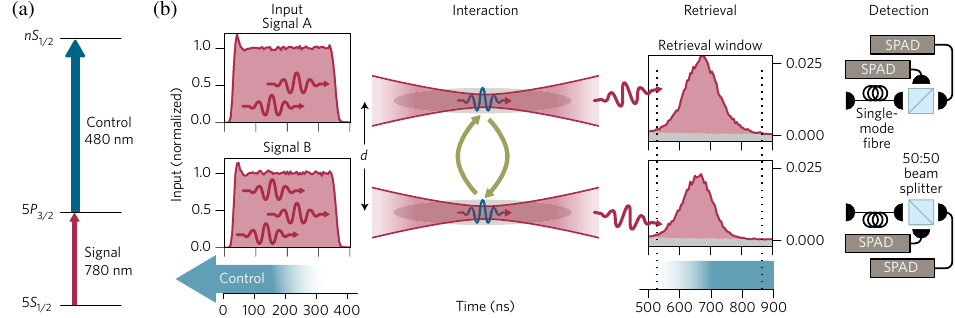}
\caption{\label{wufignonlinear}Experimental realization of contactless nonlinear optics mediated by
long-range Rydberg interactions. 
Reproduced with permission from  Busche {\it et al}., Nature Physics {\bf 13}(7), 655-658 (2017). Copyright 2017 Springer Nature.\cite{Busche2017}}
\end{figure*}

Even in the absence of a common medium, creating effective photon-photon interactions remains feasible.\cite{yang2016interacting} Busche {\it et al.}'s study delved into a novel approach to such photon interactions utilizing Rydberg interactions.\cite{Busche2017} As depicted in Fig.~\ref{wufignonlinear}, photons stored as collective Rydberg excitations in spatially separated optical media interact through vdw interactions, inducing non-uniform phase shifts that alter the retrieval modes of the photons. These interaction-induced phase shifts are pivotal in shaping the overall interaction between the photons, resulting in observable effects like reduced retrieval in the photons' initial modes and spatial anti-correlations in observed photon statistics. 
The experimental arrangement utilized highly focused signal beams and tiny optical tweezers to confine cold atoms, demonstrating the high level of precision and control in investigating non-contact nonlinear optics.
Results exhibit strong agreement between theory and experiment across varying interaction strengths and distances, indicating the feasibility of implementing contactless effective photon-photon interactions in practical photonic devices. This contactless characteristic of interactions unveils new opportunities for scalable multichannel photonic devices, exploration of strongly correlated many-body dynamics using light, and the development of advanced quantum technologies based on photon interactions mediated by Rydberg states.

\subsection{Single-photon engineering}

Rigid dipole blockade effect in Rydberg atoms allows one to attain efficient single-photon engineering, including single-photon generation and storage due to collectively enhanced strong couplings between single Rydberg superatoms and few-photon light pulses.\cite{PhysRevX.7.041010} In Ref.~\onlinecite{PhysRevA.66.065403}, Saffman and Walker conducted a detailed analysis of collective emission and proposed the protocol for realizing deterministic single-photon source using Rydberg superatom. Instead of using Rydberg EIT to store a single photon, they show that the Rydberg superatom can be coherently prepared by using two-photon excitation to drive a many-body Rabi oscillation, hence efficiently transferred into a highly directional single photon. Honer {\it et al.}\cite{PhysRevLett.107.093601} and Tresp {\it et al.}\cite{PhysRevLett.117.223001}  further examined photon interactions with a Rydberg superatom under controlled dephasing conditions and discovered that induced dephasing may dictate the absorption of a single photon from any probing field, which was recently employed in the controlled multi-photon subtraction process.\cite{stiesdal2021controlled}

\begin{figure}
\centering  
\includegraphics[width=0.9\linewidth]{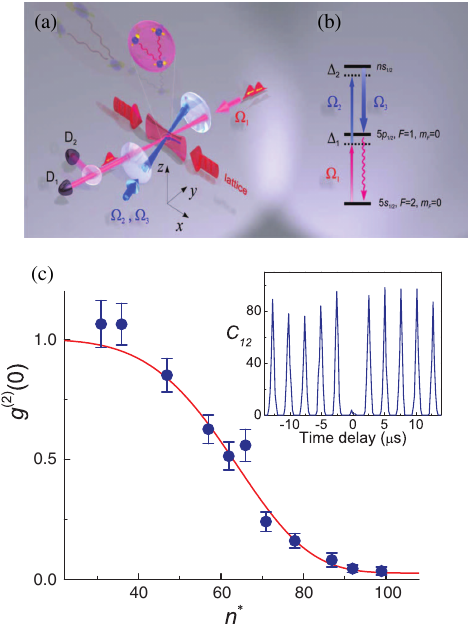}
\caption{\label{SP} Single-photon source with Rydberg superatom. (a) Illustration of experimental protocol. (b) Relevant atomic levels. (c) Measured second-order-correlation function $g^{2}(0)$ as a function of the effective principal quantum number.
Reproduced with permission from Dudin {\it et al.}, Science {\bf 336}(6083), 887–889 (2012). Copyright 2012 The American Association for the Advancement of Science.\cite{doi:10.1126/science.1217901}}
\end{figure}

A deterministic single-photon source based on Rydberg superatom was experimentally realized in 2012. \cite{doi:10.1126/science.1217901} As shown in Fig.~\ref{SP}, Dudin and Kuzmich performed 780~nm-480~nm two-photon Rydberg excitation and photon retrieval in an ensemble of $^{87}$Rb atoms confined within the blockade radius. When high-lying Rydberg states ($n\sim 100$) are used, the measured second-order-correlation function $g^{2}(0)$ of the retrieved photon approaches zero, indicating the successful preparation of a single photon. Further research in Rydberg single-photon source focused on analyzing and improving the quality of generated single photons.\cite{PhysRevLett.123.203603,Ornelas-Huerta:20, PhysRevResearch.3.033287, Shi2022} 
The state-of-the-art Rydberg single-photon sources feature near-unity purity and indistinguishability, which are comparable to those in other leading single-photon physical platforms\cite{aharonovich2016solid} and have been used to achieve a high-fidelity photonic quantum logic gate using linear optical approach.\cite{Shi2022}

\begin{figure}
\includegraphics[width=0.9\linewidth]{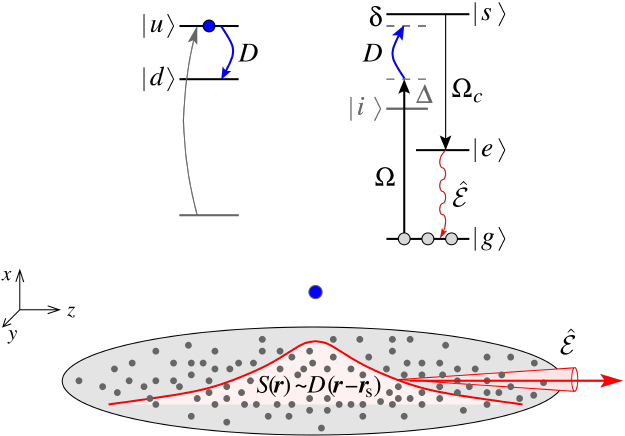}
\caption{\label{wufig5} Transition $|u\rangle\rightarrow|d\rangle$ of a source atom (top left) is coupled
nonresonantly to transition $|i\rangle\rightarrow|s\rangle$ of the medium atoms (top right) via DDE interaction $D$. A laser pulse $\Omega$ couples ground state $|g\rangle$ to Rydberg state $|i\rangle$ and leads to the transition $|g\rangle\rightarrow|s\rangle$ with the help of DDE transition $|i\rangle|u\rangle\rightarrow|s\rangle|d\rangle$. 
The resulting single Rydberg excitation exhibits the profile $S(\textbf{r})\propto D(\textbf{r}-\textbf{r}_{s})$ and can be converted into a propagating photon $\mathcal{E}$ by applying the control field $\Omega_{c}$.
Reproduced with permission from Petrosyan {\it et al}., Phys. Rev. Lett. {\bf 121}(12), 123605 (2018). Copyright 2018
American Physical Society.\cite{PhysRevLett.121.123605}}
\end{figure}

\begin{figure*}
\includegraphics[width=0.95\linewidth]{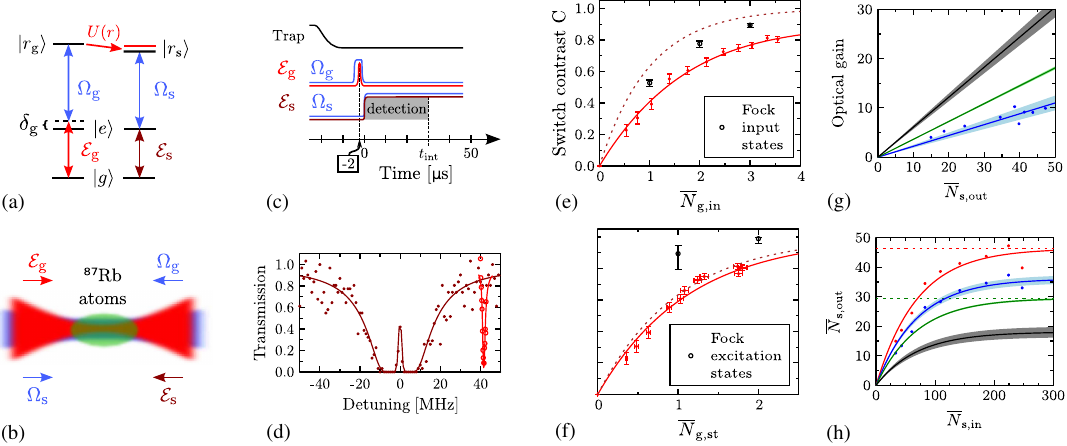}
\caption{\label{wufig6} (a) Level scheme, (b) simplified schematic, and (c) pulse sequence for the realization of an all-optical
transistor with $U(r)$ denoting the dipole-dipole interaction between two atoms of
distance $r$ and in states $|r_{g}\rangle$ and $|r_{s}\rangle$, respectively. (d) Absorption spectrum of
the source field (dots) exhibiting an EIT window at $\delta_{s}=0$ and that of the gate field (circles) around
$\delta_{g}=40$~MHz. The contrast of the switches (red solid) against mean numbers of (e) incident and (f) stored gate photons in
coherent states is presented with fundamental limits (pink dashed) set by the corresponding photon statistics. (g) Gate
photon optical gain and (h) source photon transfer function with $\overline{N}_{\rm g,in}=0$ (red) and $\overline{N}_{\rm g,in}=0.75$
(blue) for coherent states, while $N_{\rm in,Fock}=1$ (green) and $N_{\rm st,Fock}=1$ (black) for Fock states.
Reproduced with permission from
Gorniaczyk {\it et al.}, Phys. Rev. Lett. {\bf 113}(5), 053601 (2014). Copyright 2014
American Physical Society.\cite{PhysRevLett.113.053601}}
\end{figure*}

In 2018, Petrosyan and M{\o}lmer proposed a more efficient free space scheme to create
deterministic single photons in a well-defined spatio-temporal mode.\cite{PhysRevLett.121.123605} This is done by first excite
a single source atom into a high-lying Rydberg state $|u\rangle$, which couples to an ensemble of ground state atoms via
dipole-dipole exchange (DDE) interactions so that its transition to another Rydberg state $|d\rangle$ is accompanied by
the creation of a spatially extended spin wave $S(\textbf{r})$ of a single Rydberg excitation in the ensemble, as shown
in Fig.~\ref{wufig5}. This collective single-Rydberg excitation can be converted to a propagating optical photon $\mathcal{E}$ as a coherent control field $\Omega_{c}$ is switched on to form the ladder EIT configuration. This nontrivial scheme relies on neither the strong coupling of a single emitter to a resonant cavity nor the heralding of collective excitation or
complete Rydberg blockade of multiple excitations in a mesoscopic atomic ensemble. Similarly, it is feasible to achieve reliable quantum memory at the single-photon level by absorbing all extra photons inside each blockade volume of
the Rydberg superatoms.\cite{PhysRevLett.112.243601,Li2016,distante2017storing,zhang2022single}

The dipole blockade and exchange mechanisms described above used for generating single-photon sources can also be extended to design
usually unattainable single-photon devices, including {single-photon switch,\cite{PhysRevLett.112.073901}  single-photon transistor, \cite{PhysRevLett.113.053601,PhysRevLett.113.053602} and  single-photon absorber}. \cite{PhysRevLett.117.223001,PhysRevLett.107.093601}
For instance, one may store a gate light pulse containing only one photon on average as a collective Rydberg excitation in cold atoms, in the presence of which the transmission of the subsequent target pulse is largely suppressed due to dipole blockade.\cite{PhysRevLett.112.073901} Such a single-photon switch offers many interesting prospects ranging from quantum communication to quantum computing, such as heralding a successful
quantum memory event, nondestructively detecting an optical photon, and building a photonic quantum-logic gate. Moreover,
it is viable to realize an all-optical single-photon transistor by mapping gate and source (target) photons
to strongly interacting Rydberg excitations with different principal quantum numbers~\cite{PhysRevLett.113.053601} or by
suppressing and enhancing the transmission of target (source) photons through dipole blockade and F\"orster resonances,
respectively.\cite{PhysRevLett.113.053602}

To be more specific, we can see from Figs.~\ref{wufig6}(a)-\ref{wufig6}(d) that an all-optical transistor can be achieved with the following two procedures. First, store photons in a gate field $\mathcal{E}_{g}$ as collective Rydberg excitation in a sample
of cold $^{87}$Rb atoms by coupling the ground state $|g\rangle$ with the Rydberg state $|r_{g}\rangle$ through a control field
$\Omega_{g}$. Second, a source field $\mathcal{E}_{s}$ is sent through the same medium at a reduced group velocity due
to the resonant EIT provided by the control field $\Omega_{s}$ coupling to the Rydberg state $|r_{s}\rangle$. Then individual
source photons can travel with little absorption through the atomic sample in the absence of a gate excitation for a
low source photon rate or are strongly attenuated by the scattering from the intermediate state $|e\rangle$ in the blockade
regime where Rydberg interactions between states $|r_{g}\rangle$ and $|r_{s}\rangle$ result in a two-level absorbing
system. This realizes a transistor working at the single-photon level with its main performance illustrated
in Figs.~\ref{wufig6}(e)-\ref{wufig6}(h). First, we can see from Fig.~\ref{wufig6}(e) that the switch contrast $C$ plotted against the mean number of incident gate photons $\overline{N}_{g,{\rm in}}$ is clearly below the fundamental limit set by the coherent-state photon
statistics. Figure~\ref{wufig6}(f) shows, however, that $C$ plotted against $\overline{N}_{g,{\rm st}}$ (the mean number of stored gate
photons) gets much closer to the corresponding fundamental limit. It is more important to note that, on average, about $3$ input or $2$ stored gate photons are already enough to achieve a switch contrast over $0.8$. Figure~\ref{wufig6}(g) further shows
that the optical gain is up to $20\sim30$ (limited by the self-blocking effect of the source photons) where a coherent input
with $\overline{N}_{g,{\rm in}}=0.75(3)$ gate photons can be used to effectively control the transmission of $30\sim50$ source
photons. Finally, we find in Fig.~\ref{wufig6}(h) a nonlinear saturation behavior due to self-blockade in terms of the source photon transfer function from $\overline{N}_{s,{\rm in}}$ to $\overline{N}_{s,{\rm out}}$. Hence a much higher optical gain is
expected to occur provided the source photon loss due to self-blockade can be overcome.

{ Recently, Rydberg polaritons have been utilized for the generation and manipulation of quantum vortices { in an extreme regime of quantum nonlinear optics at the single-photon level}.\cite{doi:10.1126/science.adh5315} By employing EIT in a ladder-type scheme involving 100$S_{1/2}$ Rydberg state of a ultracold rubidium gas, Rydberg polaritons enable a strong effective interaction between two or three photons. This interaction leads to faster phase accumulation for co-propagating photons and the formation of quantum vortex-antivortex pairs within the two-photon wave function. Furthermore, the presence of three photons results in the emergence of vortex lines and a central vortex ring, indicating a genuine three-photon interaction. The wave function topology, influenced by two- and three-photon bound states, imposes a conditional phase shift per photon, opening prospect for deterministic quantum logic operations.}
\subsection{Hybrid Rydberg systems}
{Hybrid quantum systems are systems that integrate various types of quantum components or platforms to capitalize on their complimentary qualities. These systems seek to exploit the distinct qualities of diverse quantum technologies in order to produce more robust, scalable, and adaptable quantum devices.}
Hybrid systems involving Rydberg atoms promise intriguing possibilities for quantum information processing. It was shown
that a JC model could be realized in the optical domain for an ensemble of atoms by exploiting their
collective couplings to a quantized cavity mode.\cite{PhysRevA.82.053832,PhysRevA.95.022317} This is realized by restricting the huge state
space to the ground state $|G\rangle$ and the collective excited state {$|R\rangle$ with a single Rydberg excitation} due to {Rydberg} blockade. The main benefit is that the collectively enhanced coupling {between the emitter and the cavity mode} can be much larger than
the  decay rates, leading to the strong coupling regime of an effective JC model. Trapping some atoms in a volume
smaller than the blockade sphere close to an atom chip, it is viable to attain a hybrid system consisting of a superatom coupled
to a surface phonon polariton, which can be used to subtract individual photons from a beam of light.\cite{PhysRevA.88.043810}
Hybrid systems consisting of macroscopic mechanical oscillators coupled to a small ensemble of Rydberg atoms are also
widely investigated to achieve ground state cooling, quantum state engineering, nontrivial nonlinear dynamics, bistable
Rydberg excitation, and mechanical oscillation, etc.\cite{PhysRevA.91.023813,PhysRevA.89.011801,Carmele_2014} This depends
especially on the collectively enhanced coupling arising from the dipole blockade effect of Rydberg superatoms.\cite{PhysRevLett.121.103601,Yang:22}
The strong coupling regime may also be realized between a Rydberg atomic ensemble and propagating surface phonon
polaritons on a piezoelectric superlattice.\cite{PhysRevLett.117.103201} These hybrid Rydberg systems provide a promising
platform for extending the contents of nonlinear quantum optics and photonics, essential for developing future quantum control
techniques. Moreover, when combined with Floquet engineering techniques, polaritons, formed by photons entering a cavity and hybridized with atomic Rydberg excitations, can be efficiently manipulated to interact with each other.\cite{clark2019interacting} They can also be simulated to exhibit topological order, such as the Laughlin state.\cite{clark2020observation}

\begin{figure}
\includegraphics[width=0.9\linewidth]{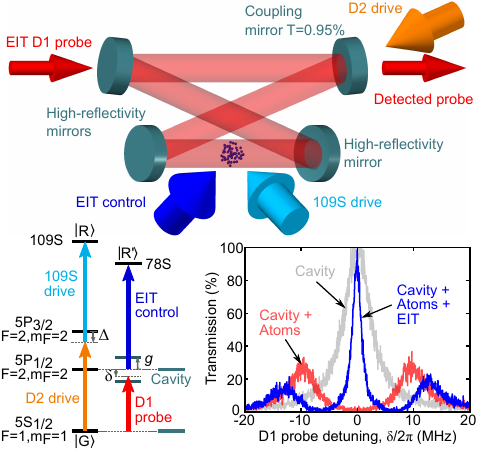}
\caption{\label{wufig8} The experimental setup involves measuring transmission, with EIT spectra depicted in blue (presence of EIT control) or red (absence of EIT control), normalized against the maximum transmission of an empty cavity (depicted in gray).
Reproduced with permission from Vaneecloo {\it et al}., Phys. Rev. X {\bf 12}(2), 021034 (2022). Copyright 2022
American Physical Society, licensed under a Creative Commons Attribution 4.0 International license.\cite{PhysRevX.12.021034}}
\end{figure}

\begin{figure}
\includegraphics[width=0.9\linewidth]{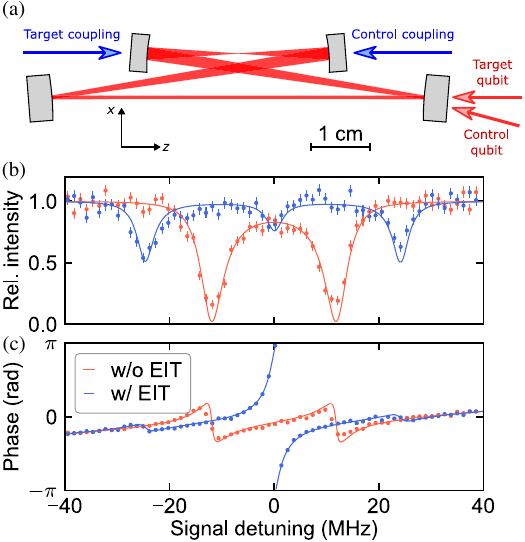}
\caption{\label{wufigadd} (a) A bow-tie cavity composed of two concave mirrors and two convex mirrors. Relative intensity (b) and (c) phase of signal light as a function of signal detuning referring to cavity-enhanced Rydberg EIT spectra. Red (blue) data and lines are attained in the absence (presence) of coupling light.
Reproduced with permission from Stolz {\it et al}., Phys. Rev. X {\bf 12}(2), 021035 (2022). Copyright 2022
American Physical Society, licensed under a Creative Commons Attribution 4.0 International license.\cite{PhysRevX.12.021035}}
\end{figure}

Below, we will explain the above discussions in more detail by focusing on recent experiments.\cite{PhysRevX.12.021034,PhysRevX.12.021035,Magro2023}
As can be seen in Fig.~\ref{wufig8}, by placing a small cloud of $^{87}$Rb atoms inside a cavity {with a finesse of} $\mathcal{F}=590$,
one observes the vacuum Rabi splitting of two transmission peaks located at $\delta/2\pi=\pm10$~MHz, indicating a cavity-mode coupling constant ($g/2\pi=10$~MHz) exceeding both {the} cavity decay rate ($\kappa/2\pi=2.9$~MHz) and {the} atomic decay rate ($\gamma/2\pi=3$~MHz). Exciting these atoms to a high Rydberg state $|R^{\prime}\rangle$ by the EIT control and D1 probe beams, a narrower EIT window
is found to open on resonance corresponding to a dark polariton state mixing a photon in the cavity and a Rydberg excitation in the atomic cloud. On the other hand, the $109S$ drive and D2 drive beams can drive these atoms to another high Rydberg state $|R\rangle$ via a two-photon Raman resonance so that they behave as a two-level superatom due to the blockade of multiple Rydberg excitations. {Atoms in states $|R\rangle$ and $|R^{\prime}\rangle$ further strongly interact  with another pair of Rydberg states through F\"orster resonance.} Then, by shifting the state $|R\rangle$ out of resonance, a superatom in the state $|R^{\prime}\rangle$ destroys the EIT response to the D1 probe, which drives the photonic component of the $|R^{\prime}\rangle$ polariton. This scheme has been exploited to realize the high-efficiency single shot optical detection of a Rydberg superatom, the superatom state-dependent $\pi$ phase shift of light reflected by the cavity, and the deterministic preparation of freely propagating photonic qubits with negative Wigner functions,\cite{PhysRevX.12.021034,Magro2023} by taking advantage of two complementary approaches: i.e., cavity quantum electrodynamics and interacting atomic ensembles.\cite{Kumar_2016}

It is also worth noting that a bow-tie cavity has been incorporated into a Rydberg EIT experiment for demonstrating a polarization CNOT gate between two optical pulses interacting with hundreds of trapped $^{87}$Rb atoms at an average efficiency of about $41.7\%$, far exceeding that in previous experiments.\cite{PhysRevX.12.021035} This atom-cavity system in a ring geometry minimizes dephasing resulting from the combination photon recoil of atomic motion and is characterized by $(g,\kappa,\gamma)=2\pi\times(1.0,2.3,3.0)$~MHz, thus yielding finesse $\mathcal{F}=350$ and a collective cooperativity of $21$ facilitating the enhanced interaction of single photons and Rydberg atoms. The four-mirror bow-tie cavity is shown in Fig.~\ref{wufigadd}(a) where the target and control coupling beams are guided into the cavity from different concave mirrors, while the target and control qubits are input from a common convex mirror. This ring resonator is necessary for converting the $\pi$ conditional phase shift into a two-photon gate for polarization qubits. The relative intensity and phase of the target signal reflected from the cavity have been measured as a function of the signal detuning from the cavity resonance in Figs.~\ref{wufigadd}(b) and \ref{wufigadd}(c), respectively. In the absence of an EIT coupling light, the intensity exhibits a normal-mode splitting, while the phase shows two dispersive features, one at a normal mode (see red data). As the EIT coupling light is on, a narrow EIT feature appears at resonance so that a $2\pi$ phase boost may occur as the signal detuning is tuned through zero (see blue data). This forms the basis for a $\pi$ conditional phase shift, essential for constructing a CNOT gate.

Another interesting hybrid setup involves polar molecules and Rydberg atoms trapped in optical tweezer arrays in which the Rydberg blockade due to the charge-dipole interaction is demonstrated. \cite{PhysRevLett.131.013401} This setup is highly beneficial because quantum information can be encoded in the internal states of the molecule, and quantum gates can be performed by utilizing the strong Rydberg-Rydberg interactions of atoms, with little sensitivity to the motional states of the particles. \cite{C1CP21476D, PRXQuantum.3.030339, PRXQuantum.3.030340}

\section{CONCLUSION AND OUTLOOK}
\label{concl}

In conclusion, Rydberg atoms have emerged as one of the most promising platforms for probing quantum physics and quantum technological applications. The essential elements necessary for understanding Rydberg physics are covered in the first part of this review article, particularly the optical excitation of Rydberg states and the emergence of different types of Rydberg-Rydberg interactions, such as vdw and dipolar interactions, including the tunability using external fields. The strong interatomic interactions lead to the emergence of superatoms in Rydberg gases via the Rydberg blockade. When driving by resonant lasers, many-body Rabi oscillations can be found. The superatom approach is an efficient way to describe the dynamics and collective states, while the underlying mechanism can be used to generate robust and scalable entanglement in neutral atoms. Further, the same mechanism can create the entanglement between atoms and photons and between two photons. In an atom array, Rydberg blockade and superatoms serve a mechanism to emulate many-body spin models to create various many-body states and intricate quantum dynamics.

The remarkable collective excitation properties of Rydberg atoms hold immense promise for advancing quantum information processing and quantum optics. It is expected that an increasing number of experimental achievements and theoretical proposals will be seen in the coming years. Future developments have a number of exciting perspectives.  Quantum simulation, quantum computation and quantum optics applications are promising growing area. To fully unlock the potential of Rydberg atom systems, new schemes and experimental techniques are needed in order to mitigate limitations (such as spin wave dephasing) in existing experiments. Moreover, by combining different quantum systems, their individual properties can be leveraged in order to realize unique functionality. Recently, hybrid systems in which Rydberg excitations combine with Bose-Einstein condensates, \cite{PhysRevLett.100.033601, PhysRevA.103.063307} trapped ions, \cite{zhang2020submicrosecond,MOKHBERI2020233} polar molecules,~\cite{PhysRevLett.131.013401,PRXQuantum.3.030340,PRXQuantum.3.030339} mechanical oscillators,~\cite{PhysRevLett.113.023601} etc., pave the way for the exploration of novel applications and physics in the future.

\section*{ACKNOWLEDGMENTS}
The authors extend their sincere gratitude to the anonymous reviewers for their thorough reading and valuable suggestions, which have significantly improved the readability of the manuscript. Additionally, we would like to thank F. Cesa,  C. S. Adams, S. D\"{u}rr, S. Hofferberth, Y. O. Dudin, D. Petrosyan, J. Zeiher, and A. Ourjoumtsev for their valuable comments. This work was supported by the National Natural Science
Foundation of China (NSFC) under Grant Nos. 12174048, 12274376, 12074061, 12374329, and U21A6006, and the National Key Research and Development Program of China under Grant No. 2021YFA1402003.
R.N. acknowledges DST-SERB for Swarnajayanti fellowship File No. SB/SJF/2020-21/19, the MATRICS grant (MTR/2022/000454) from SERB, and funding from National Mission on Interdisciplinary Cyber-Physical Systems (NM-ICPS) of the Department of Science and Technology, Govt. of India through the I-HUB Quantum Technology Foundation, Pune INDIA. W.L. acknowledges support from the EPSRC through Grant No.
EP/W015641/1 and the Going Global Partnerships Programme of the British Council (Contract No. IND/CONT/G/22-23/26).

\vspace{12pt}
\section*{AUTHOR DECLARATIONS}
\subsection*{Conflict of Interest}
The authors have no conflicts to disclose.

\vspace{12pt}
\subsection*{Author Contributions}
{\bf Xiao-Qiang Shao}: Project administration (equal); Funding acquisition (equal); 
 Writing-original draft (equal); Writing-review \&
editing (equal). {\bf Shi-Lei Su}: Funding acquisition (equal); 
 Writing-original draft (equal); Writing-review \&
editing (equal). {\bf Lin Li}: Funding acquisition (equal); 
 Writing-original draft (equal); Writing-review \&
editing (equal). 
 {\bf Rejish Nath}: Funding acquisition (equal); 
 Writing-original draft (equal); Writing-review \&
editing (equal).
 {\bf Jin-Hui Wu}: Funding acquisition (equal); 
 Writing-original draft (equal); Writing-review \&
editing (equal).
 {\bf Weibin Li}: Project administration (equal); Funding acquisition (equal); 
 Writing-original draft (equal); Writing-review \&
editing (equal).

\vspace{12pt}
\section*{DATA AVAILABILITY}
The data of Figs.~\ref{scaling} and \ref{LLfig5} generated in this review article are available from the corresponding authors upon reasonable request.

\section*{REFERENCES}

\bibliography{APR-share.bbl}

\end{document}